\documentclass[11pt]{article}

\usepackage{lineno,hyperref,bm}
\usepackage[utf8]{inputenc}
\usepackage[margin=1.3in]{geometry}
\usepackage{bm}
\usepackage{comment}
\usepackage{lscape}
\usepackage{float,enumitem}
\usepackage{amsmath,amssymb,amsthm,color,CJK,mathrsfs}
\usepackage[title]{appendix}
\usepackage{graphicx}
\usepackage{subcaption}
\usepackage{authblk}

\DeclareMathOperator*{\argmin}{arg\,min}
\DeclareMathOperator*{\argmax}{arg\,max}
\usepackage{authblk}

\def\keywordfont{\fontsize{10}{11}\selectfont}
\newenvironment{keywords}{\par\addvspace{8pt}%
	\keywordfont\noindent{\textbf{Keywords}}:\ \ignorespaces%
}{\par\addvspace{23pt}}

\date{}
\begin{document}


\title{Tangent functional canonical correlation analysis for densities and shapes, with applications to multimodal imaging data}

\author{Min Ho Cho \footnote{Department of Applied and Computational Mathematics and Statistics, The University of Notre Dame} \ \ Sebastian Kurtek \footnote{Department of Statistics, The Ohio State University}  \ \ Karthik Bharath 
\footnote{School of Mathematical Sciences, The University of Nottingham}
}


\maketitle

\begin{abstract}
It is quite common for functional data arising from imaging data to assume values in infinite-dimensional manifolds. Uncovering associations between two or more such nonlinear functional data extracted from the same object across medical imaging modalities can assist development of personalized treatment strategies. We propose a method for canonical correlation analysis between paired probability densities or shapes of closed planar curves, routinely used in biomedical studies, which combines a convenient linearization and dimension reduction of the data using tangent space coordinates. Leveraging the fact that the corresponding manifolds are submanifolds of unit Hilbert spheres, we describe how finite-dimensional representations of the functional data objects can be easily computed, which then facilitates use of standard multivariate canonical correlation analysis methods. We further construct and visualize canonical variate directions directly on the space of densities or shapes. Utility of the method is demonstrated through numerical simulations and performance on a magnetic resonance imaging dataset of Glioblastoma Multiforme brain tumors.
\end{abstract}
\begin{keywords}
Hilbert manifold; Square-root transform; Elastic shape analysis; Intrinsic principal component analysis.
\end{keywords}


\section{Introduction}
\label{sec:intro}

Canonical correlation analysis (CCA) is a statistical technique used to investigate relationships between two sets of variables; it was first introduced by Hotelling \cite{hotelling1936relations}. Classical CCA in the multivariate setting seeks linear combinations of two sets of variables, also called canonical variates, which maximize correlation \cite{mardia1979multivariate}. Extensions of classical CCA to the functional data setting have also been previously considered in the literature. The main challenge arises from the infinite-dimensionality of the resulting representation space. Leurgans et al.~\cite{leurgans1993canonical} were the first to propose an extension of CCA for functional data analysis (FDA). They showed that regularization is essential when estimating the canonical variates and imposed it via smoothing. Subsequently, multiple alternate versions of functional CCA (FCCA) have been proposed that utilize various forms of regularization  \cite{ramsay1997functional}, e.g., the kernel approach in \cite{he2004methods}. Finally, Shin and Lee \cite{shin2015canonical} developed FCCA for irregularly and sparsely observed functional data. He et al.~\cite{he2004methods} review four different computational methods for estimating canonical correlations and canonical variates in the context of FDA. More generally, for a comprehensive introduction to recent advances in FDA, we refer the interested reader to recent surveys by Aneiros et al. \cite{aneiros2019recent} and Goia and Vieu \cite{DBLP:journals/ma/GoiaV16}.

All of the aforementioned methods assume that the representation space of functions is Euclidean, and in particular a Hilbert space. In many applications, however, one is interested in modeling non-Euclidean functional data such as probability density functions (PDFs) and shapes of curves. These data objects naturally arise in various domains including medical imaging, biology and computer vision. PDFs are non-negative functions that integrate to one; as a result, their representation space is an infinite-dimensional simplex. Similarly, shapes of parameterized curves lie on a quotient space of an infinite-dimensional nonlinear space due to invariance requirements, i.e., the notion of shape is invariant to translation, scale, rotation and re-parameterization. In such examples, standard FCCA methods are inappropriate to study associations between such objects.

\subsection{Motivating application: multimodal Magnetic Resonance Imaging}
\label{sec:motapp}

\begin{figure}[!t]
	\centering
	\includegraphics[scale=0.3]{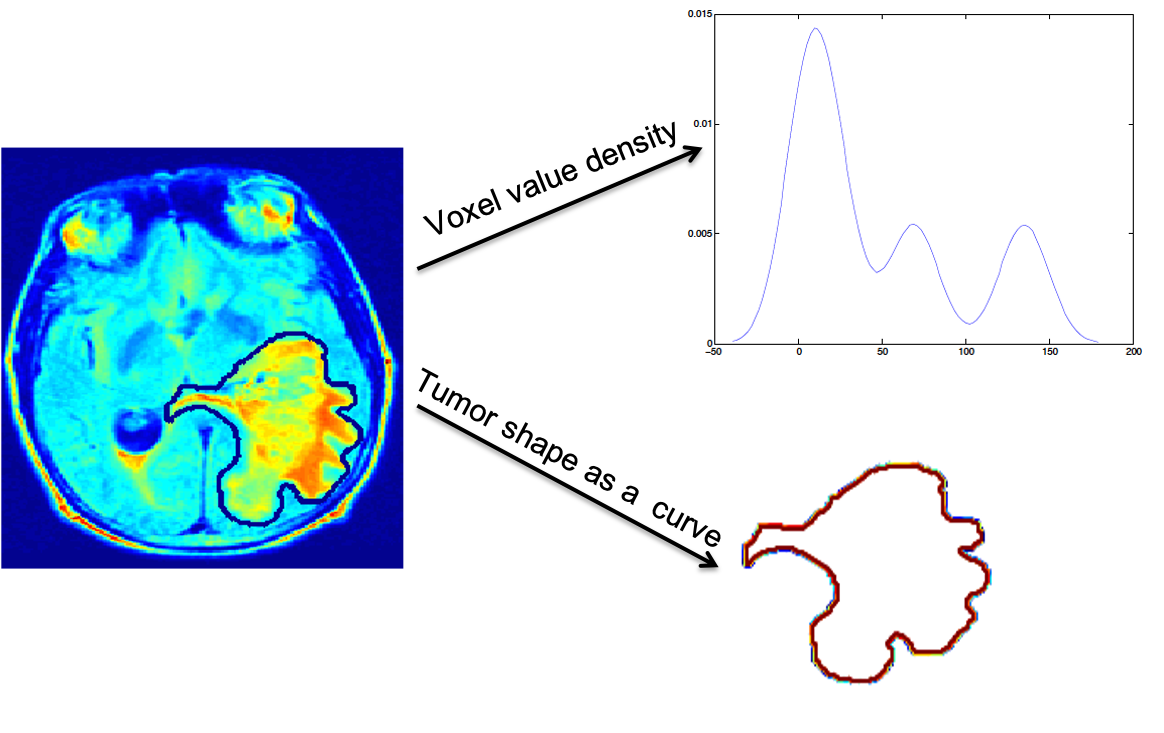}
	\vspace{-.6cm}
	\caption{Two non-Euclidean functional data objects extracted from an MRI image of a patient diagnosed with GBM.}
	\label{fig:GBM}
\end{figure}

Glioblastoma multiforme (GBM) is the most common malignant brain tumor in adults \cite{holland2000} with very poor prognosis; the median survival time for patients diagnosed with GBM is approximately 12 months \cite{tutt2011}. The most common noninvasive technique for assessing GBM progression and tumor heterogeneity is via magnetic resonance imaging (MRI) \cite{just2014}. In terms of heterogeneity, both the textural appearance of the tumor in the image as well as its geometry are informative descriptors. The textural characteristics of the tumor can be summarized via the probability density function (PDF) of intensity values inside the tumor \cite{saha2016demarcate}. On the other hand, tumor geometry can be quantified via the shape of its outer contour \cite{bharath2018radiologic}. Fig. \ref{fig:GBM} offers an illustration. Associations between the two functional descriptors, which reside in non-Euclidean function spaces, can be nonlinear and complicated, e.g., a PDF of voxel values indicates clustered tumor texture which may be associated with a tumor contour whose shape deviates substantially from an ellipse; they can also be difficult to interpret. Additionally, MRI provides a wide range of imaging contrasts through multimodal images, e.g., pre-surgical T1-weighted post contrast and T2-weighted fluid-attenuated inversion recovery. Each of these sequences identifies different types of tissue and displays them using varying contrasts based on the tissue characteristics. From each modality, a PDF and a tumor contour can be extracted, and associations between PDFs (tumor contours) between modalities are also informative descriptors of tumor heterogeneity, but, as mentioned above, can be difficult to interpret.

What is thus of practical importance is a version of FCCA on a locally linear representation of two non-Euclidean functional data objects, which track infinitesimal linear associations between the two. In particular, we consider: (i) pairs of voxel value PDFs corresponding to two different MRI modalities, (ii) pairs of tumor shapes corresponding to two different MRI modalities, and (iii) pairs of voxel value PDFs and shapes extracted from the same MRI modality. 

\subsection{Contributions}

We consider a Riemannian geometric setting for the non-Euclidean spaces of PDFs and shape curves, and utilize the tangent space parameterization of the resulting manifolds at a fixed point. The locally linear representations of PDFs and shape curves are obtained by mapping them onto the tangent space of a fixed point, which will be taken to be the Karcher mean with respect to the Riemannian metric. This approach is sensible when the variability in the functional data objects is not large and the data is fairly concentrated.

For the space of PDFs, we adopt the Fisher-Rao Riemannian metric; the metric has the attractive property that under a simple square root transformation \cite{bhattacharyya1943measure} it simplifies to the standard $\mathbb{L}^2$ metric \cite{rao1945information,kass1997geometrical,srivastava2007riemannian}, and the representation space of PDFs is mapped to the positive orthant of the unit Hilbert sphere. This facilitates efficient analytic computation of all relevant geometric quantities. Correspondingly, we adopt the elastic shape analysis framework to study shapes of curves \cite{srivastava2016functional}. In this approach, shape changes are measured via an elastic Riemannian metric \cite{younes1998computable,mio2007shape} that captures the amount of stretching and bending needed to deform one shape into another. As with PDFs, a simplifying transformation, called the square root velocity function \cite{srivastava2011shape}, is used to facilitate computation of geometric quantities on the elastic shape space which, again, is a quotient space of a submanifold of an appropriate Hilbert sphere. A major benefit of this framework is that it allows parameterization-invariant statistical analysis of shapes. 

A well-known feature of FCCA is the need for smoothing or dimension reduction in order to circumvent arbitrarily high correlations between canonical variates \cite{leurgans1993canonical}. Accordingly, our definition of FCCA for PDFs and shape curves combines the local linearization in tangent spaces with dimension reduction using a data-driven basis for the chosen tangent space; this basis is obtained by an eigen decomposition of the (empirical) covariance operator defined using the locally linearized functional data objects. Dimension reduction is achieved by projecting them onto a finite-dimensional subspace spanned by a finite number of eigenbasis functions. The resulting coefficient vectors represent finite-dimensional Euclidean representations of PDFs and shape curves, and can then be used to carry out standard multivariate CCA. 

Since each element of the coefficient vector is associated with a basis element, the proposed tangent FCCA (TFCCA) approach enables us to easily visualize a particular direction specified by a canonical variate by first computing the associated vector in the tangent space and then projecting it back on the manifold of PDFs or shape curves. These canonical variate directions maximize the correlation between the given pairs of PDFs and/or shapes. Moreover, the canonical variates can be used as predictors in a regression model when a response is available \cite{luo2016canonical}. Along the lines of how existence of FCCA for random functions in  Hilbert spaces is established, we examine conditions required for existence of FCCA on a chosen tangent space and explicate influence of the choice. As such, our work represents a novel contribution to the FCCA literature by considering non-Euclidean functional data in the form of PDFs and shape curves, and utilizing the Riemannian geometry of their representation spaces to define suitable FCCA procedures.

\subsection{Notation and paper organization} 

Vectors/matrices and sequences are denoted in bold while all functional data objects are given as normal text. We use $\|\cdot\|$ and $\langle\langle\cdot,\cdot\rangle\rangle$ to denote the functional $\mathbb{L}^2$ norm and inner product, respectively; non-standard Riemannian metrics are denoted by $\langle\langle\langle\cdot,\cdot\rangle\rangle\rangle$. The standard Euclidean norm and inner product in $\mathbb{R}^n$ are denoted by $|\cdot|$ and $\langle\cdot,\cdot\rangle$, respectively.

The rest of this paper is organized as follows. In Section \ref{sec:frame}, we review the relevant Riemannian geometric frameworks for PDFs and shapes that are used to compute local linearizations in tangent spaces and to carry out dimension reduction. Section \ref{sec:tcca} examines existence of FCCA on tangent spaces, and describes the proposed FCCA approach for paired sets of PDFs and/or shapes. We also overview the framework of canonical variate regression.
In Section \ref{sec:simul}, we assess the proposed methods using a simulation study for PDFs. In Section \ref{sec:empirical}, we apply our framework to study associations between textural and shape features of GBM brain tumors in two MRI modalities. We close with a brief discussion and some directions for future work in Section \ref{sec:conclusion}. The supplement contains (i) a simulation study for shape curves, (ii) TFCCA results for PDFs and shapes estimated from the same MRI modality, and (iii) visualizations of canonical variate regression-based directions.


\section{Geometric framework}
\label{sec:frame}

We begin by reviewing details of the Fisher-Rao (F-R) Riemannian framework for analyzing PDFs, and elastic shape analysis. For brevity, we do not provide all details of these two frameworks, and refer the reader to Srivastava and Klassen \cite{srivastava2016functional}. Saha et al. \cite{saha2016demarcate} also provide a comprehensive summary of the F-R framework for PDFs. The elastic shape analysis framework is implemented in the \texttt{fdasrvf} package in R.

\subsection{Finite-dimensional tangent coordinates for PDFs}
\label{sec:frame_pdf}
We consider univariate PDFs on an interval of $\mathbb R$ which, without loss of generality, is assumed to be $[0,1]$. 
The set $\mathcal{P} = \big\{p : [0,1] \rightarrow \mathbb{R}_{>0}~\big|~\int^1_0 p(t) dt = 1 \big\}$ of PDFs is a Banach manifold with a tangent space at a point $p\in\mathcal{P}$ given by $T_p(\mathcal P)=\{\delta p: [0,1]\rightarrow\mathbb{R}~\big|~\int_0^1 \delta p(t)dt=0\}$. Given $\delta p_1,\ \delta p_2 \in T_p(\mathcal P)$ the (nonparametric) F-R metric is defined as $\langle\langle\langle \delta p_1,\delta p_2\rangle\rangle\rangle_p=\int_0^1 \delta p_1(t)\delta p_2(t)\frac{1}{p(t)}dt$ \cite{rao1945information,kass1997geometrical}. 
A key property of this metric is that it is invariant to re-parameterizations of PDFs, i.e., one-to-one smooth mappings of the domain to itself \cite{cencov2000statistical}. However, the computation of various geometric quantities of interest, e.g., geodesics, exponential and inverse-exponential maps, under the F-R metric is difficult in practice since the metric changes from point to point on the space of PDFs.

To simplify the F-R metric and the representation space of PDFs, we utilize the square root transformation (SRT), defined as $\psi=+\sqrt{p}$, that was proposed by Bhattacharyya \cite{bhattacharyya1943measure}. Under this transformation, the representation space of PDFs $\mathcal P$ transforms to the positive orthant of the unit Hilbert sphere $S^\infty ([0,1])$ in $\mathbb{L}^2\big([0,1],\mathbb{R}\big)$, $\Psi=\{\psi: [0,1]\to \mathbb{R}_{> 0}\ |\ \|\psi\|^2=\int_0^1\psi(t)^2 dt=1\}$, and the F-R metric reduces to the standard $\mathbb{L}^2$ metric, i.e., given two tangent vectors $\delta\psi_1,\ \delta\psi_2\in T_\psi(\Psi)=\{\delta\psi:[0,1]\to\mathbb{R}\ |\ \langle\langle\delta\psi,\psi\rangle\rangle=\int_0^1\delta\psi(t)\psi(t)dt=0\}$, the $\mathbb{L}^2$ Riemannian metric can be computed using $\langle\langle\delta\psi_1,\delta\psi_2\rangle\rangle=\int_0^1\delta\psi_1(t)\delta\psi_2(t)dt$. 

On the Hilbert sphere $S^\infty ([0,1])$, the exponential map is well-defined on the entire tangent space \cite{billioti2017riemannian}, and so is the inverse-exponential when restricted to the positive orthant since it is a geodesically convex submanifold. Thus, the inverse-exponential map at a point $\psi^*$ represents a convenient mechanism to obtain local linear versions of points on $\Psi$, which represent infinitesimal perturbations of the points along geodesics that pass through $\psi^*$. Moreover, the simple $\mathbb L^2$ geometry under the SRT representation ensures that important geometric quantities of interest can be computed in closed-form:
\begin{itemize}
	\item Geodesic distance (for $\psi_1,\ \psi_2\in\Psi$): $d(\psi_1,\psi_2)=\cos^{-1}(\langle\langle\psi_1,\psi_2\rangle\rangle)$, i.e., the length of the great circle connecting $\psi_1$ and $\psi_2$ on $\Psi$;
	\item Exponential map $\exp_\psi: T_\psi(\Psi)\to\Psi$ (for $\psi\in\Psi$,\  $\delta\psi\in T_\psi(\Psi)$): $\exp_\psi(\delta\psi) = \cos(\|\delta\psi\|)\psi + \sin(\|\delta\psi\|) \frac{\delta\psi}{\|\delta\psi\|}$;
	\item Inverse-exponential map $\exp^{-1}_{\psi_1}: \Psi\to T_\psi(\Psi)$ (for $\psi_1,\ \psi_2\in\Psi$): $\exp^{-1}_{\psi_1}(\psi_2) = \frac{d(\psi_1,\psi_2)}{\sin (d(\psi_1,\psi_2))} \big(\psi_2 - \cos (d(\psi_1,\psi_2)) \psi_1 \big)$.
\end{itemize}

\noindent The choice of $\psi^*$ is important, and a natural data-driven choice is a sample mean SRT $\bar\psi$: given points $\psi_1,\ldots,\psi_n$, the Karcher mean $\bar \psi$ is defined as a minimizer of the variance functional $\Psi \ni\psi \mapsto \frac{1}{n}\sum_{i = 1}^{n} d(\psi, \psi_i)^2$. An algorithm for calculating $\bar\psi$ is presented in Appendix A1 in \cite{saha2016demarcate}.

Next, dimension reduction of the linearized PDFs is achieved through intrinsic functional principal component analysis (FPCA) on $\Psi$ with respect to the Karcher mean $\bar \psi$. For $i = 1, \dots, n$, we first compute $\delta\psi_i = \exp^{-1}_{\bar{\psi}}(\psi_i)$ using the inverse-exponential map. The sample covariance function $k_n(s,t)=\frac{1}{n-1} \sum_{i=1}^n \delta \psi_i(s)\delta \psi_i(t)$ can be used as the kernel of the covariance operator $\mathcal K_n \delta \psi_i(t)=\int_0^1 k_n(s,t)\delta \psi_i(s)ds$. The sample eigenfunctions $e_{n,j}$ and eigenvalues $\lambda_{n,j}$  of $\mathcal K_n$ are then estimated as solutions to $\mathcal K_n e_{n,j}(t)=\lambda_{n,j} e_{n,j}(t),\ j=1,\ldots,n$. In practice, consider the $N \times 1$ vector $\bm{\delta\psi}_i$ obtained by sampling $\delta \psi_i$ using $N$ points in $[0,1]$. The singular value decomposition (SVD) of the sample covariance matrix, of size $N \times N$, is given by $\frac{1}{n-1}\sum_{i=1}^{n}\bm{\delta\psi}_i \bm{\delta\psi}_i^T=\bm U \bm \Sigma \bm U^T$, where $\bm U$ contains the matrix of discretized eigenfunctions $e_{n,j}$ as its columns and $\bm \Sigma$ is a diagonal matrix containing the principal component variances ordered from largest to smallest. In most cases, $n<<N$, and thus, this matrix is not full rank. Finally, the tangent Euclidean finite-dimensional coordinates of an SRT form an $r$-dimensional vector of PC coefficients $\bm c_i=\bm U_r^T\bm{\delta\psi}_i$, where $\bm U_r$ is the matrix containing the first $r$ columns of $\bm U$. 

To summarize, starting with a sample of PDFs $p_1,\ldots,p_n$, the sequence of operations used to obtain the requisite coordinates is
\begin{equation}
	\label{eq:pdf_steps}
	\{p_i\} \overset{SRT}\longrightarrow\{\psi_i\} \thickspace\overset{\text{mean}}\longrightarrow \thickspace\bar \psi \thickspace\overset{\exp^{-1}_{\bar \psi}(\psi_i)}\longrightarrow \thickspace\{\delta \psi_i\} \overset {\text{FPCA on }T_{\bar \psi}(\Psi)}\longrightarrow \{\bm c_i \in \mathbb R^r\} \thinspace.
\end{equation}

\subsection{Finite-dimensional tangent coordinates for shape curves}
\label{sec:frame_shape}

Let $\beta: \mathbb{S}^1 \rightarrow \mathbb{R}^2$ denote an absolutely continuous, parameterized, closed planar curve. Its shape, or the shape curve, is defined to be the equivalence class of curves $[\beta]=\{\sigma \bm O (\beta \circ \gamma) + \bm a|\sigma>0,\ \bm O \in SO(2),\ \gamma \in \Gamma,\ \bm a \in \mathbb R^2\}$ that are related to $\beta$ through a (uniform) scaling, rotation, re-parameterization and translation of $\beta$; here, $SO(2)=\big\{\bm O\in\mathbb{R}^{2\times 2}~\big|~\bm O^T\bm O=\bm O\bm O^T=I,~\det(\bm O)=+1 \big\}$ is the rotation group and 
$\Gamma = \big\{ \gamma: \mathbb{S}^1 \rightarrow \mathbb{S}^1~\big|~ 0<\dot\gamma<\infty \big\}$, where $\dot\gamma$ is the derivative of $\gamma$, is the re-parameterization group with action given by composition $(\beta,\gamma) = \beta \circ \gamma$. 

We need an appropriate distance between two shape curves $[\beta_1]$ and $[\beta_2]$. 
It is well-known that the standard $\mathbb{L}^2$ distance is not invariant to re-parameterizations \cite{srivastava2016functional}.
Mio et al. \cite{mio2007shape} and Younes \cite{younes1998computable} defined a family of first order Riemannian metrics, also referred to as elastic metrics, that are invariant to all of the aforementioned shape preserving transformations; the elastic metrics are closely related to the F-R metric for analyzing PDFs \cite{srivastava2016functional}. These metrics also provide a natural interpretation of shape deformations in terms of their bending and stretching/compression. Despite these nice mathematical properties, this family of metrics is difficult to use in practice for similar reasons as explained in the previous section. To overcome these difficulties, Srivastava et al. \cite{srivastava2011shape} introduced the square root velocity function (SRVF) $q : \mathbb{S}^1 \rightarrow \mathbb{R}^2$, defined as $q = \dot{\beta}/\sqrt{|\dot{\beta}|}$ (if $\dot\beta(t)=0$, $q(t)=0$), where $\dot\beta$ is the derivative of $\beta$. The SRVF simplifies the elastic metric to the flat $\mathbb{L}^2$ metric \cite{srivastava2011shape}, thereby facilitating easy computation. Since we suppose $\beta$ is absolutely continuous, its SRVF $q$ is square-integrable, i.e., an element of $\mathbb{L}^2\big(\mathbb{S}^1,\mathbb{R}^2\big)$. We can identify $\mathbb S^1 \cong \mathbb R /2\pi \mathbb Z \cong [0,1]$ with the unit interval, and any integral over $\mathbb S^1$ is then an integral over $[0,1]$ (assuming the starting point on $\mathbb{S}^1$ is known). One can then uniquely recover the curve $\beta$ from its SRVF $q$, up to a translation, via the relation $\beta(t)=\int_0^t q(s)|q(s)|ds$.

The SRVF representation $q$ is invariant to translations of the curve $\beta$ since it is based on the derivative. If we impose an additional unit length constraint on the curves, then the representation space of SRVFs becomes $S^\infty(\mathbb S^1) = \big \{ q: \mathbb{S}^1 \rightarrow \mathbb{R}^2~ \big|\ \|q\|^2 = \int_{\mathbb{S}_1}|q(t)|^2 = 1 \big \}$, i.e., the unit Hilbert sphere in $\mathbb{L}^2\big(\mathbb{S}^1,\mathbb{R}^2\big)$. The closed curve $\beta$ satisfies $\int _0^1 \dot{\beta}(t)dt=0$ and this leads to a corresponding closure condition under the SRVF representation $\int_0^1 q(t)|q(t)|dt=0$. 
The space $\mathcal C=\Big\{q \in S^\infty(\mathbb S^1)| \int_0^1 q(t)|q(t)|dt=0\Big\}$ is referred to as the pre-shape space, since shape preserving actions of $SO(2)$ and $\Gamma$ are yet to be accounted for. The pre-shape space $\mathcal C$ is a submanifold of $\mathbb{L}^2\big(\mathbb{S}^1,\mathbb{R}^2)$ \cite{srivastava2016functional} and the distance (induced from $S^\infty(\mathbb S^1)$) between any two SRVFs $q_1, q_2 \in \mathcal{C}$ is given by $d_{\mathcal{C}}(q_1,q_2) = \cos^{-1}\big(\langle\langle q_1,q_2 \rangle\rangle \big)$, where $\langle\langle q_1,q_2 \rangle\rangle = \int_{\mathbb{S}^1} \langle q_1(t), q_2(t)\rangle dt$.

The SRVF representation of a rotated curve $\bm O\beta$ is $\bm Oq$, while the SRVF of a re-parameterized curve $\beta\circ\gamma$ is $\big(q,\gamma\big)=\big(q\circ\gamma\big)\sqrt{\dot{\gamma}}$. For every $(\bm O,\gamma) \in SO(2) \times \Gamma$, $d_{\mathcal{C}}(\bm O(q_1,\gamma),\bm O(q_2,\gamma))=d_\mathcal{C}(q_1,q_2)$, and hence it is invariant to the product action of the rotation and re-parameterization groups. Using this result, a distance on the shape space, $\mathcal{S} = \mathcal{C} \big/ \big(SO(2)\times\Gamma\big) = \big \{ [q] ~\big|~ q \in \mathcal{C} \big \}$, can be defined as $d_{\mathcal{S}}([q_1],[q_2])= \cos^{-1} \Big( \big\langle\big\langle q_1, q_2^* \big\rangle\big\rangle \Big)$, where $q_2^*=(\bm O^*q_2,\gamma^*)$ and $(\bm O^*,\gamma^*)=\inf_{\bm O \in SO(2), \gamma \in \Gamma} \|q_1 - \bm O(q_2, \gamma)\|^2$. The quantities $\bm O^*$ and $\gamma^*$ represent the optimal global rotation and re-parameterization of $q_2$ for alignment or registration with respect to $q_1$. We use Procrustes analysis to compute $\bm O^*$, and a combination of Dynamic Programming and an exhaustive seed search to compute $\gamma^*$ \cite{srivastava2016functional}. 

Viewing $(\mathcal S, d_\mathcal S)$ as a metric space, a sample Karcher mean can be defined and computed. Let $\beta_1, \ldots, \beta_n$ be a sample of curves and $q_1, \ldots, q_n \in \mathcal{C}$ their corresponding SRVFs. 
The sample mean shape $[\bar{q}]$ is defined as the minimizer of $\mathcal S \ni [q] \mapsto \frac{1}{n}\sum^n_{i=1} d_{\mathcal{S}}([q],[q_i])^2$, and as a representative shape a single element $\bar{q}\in[\bar{q}]$ from the mean shape is chosen for subsequent analysis. A detailed algorithm for computing a mean shape is provided in \cite{kurtekcviu}.

In contrast to the framework for analyzing PDFs, two inter-related issues arise when attempting to obtain tangent coordinates for shape curves: (i) the shape space $\mathcal S$ is not a manifold \cite{srivastava2016functional} and it is hence not possible to impose a Riemannian structure on $\mathcal S$, and (ii) the tangent space at a shape $[q]$ cannot be identified with a normal subspace to $\mathcal C$ at $q$ in  $\mathbb{L}^2\big(\mathbb{S}^1,\mathbb{R}^2)$, and additionally an inverse-exponential map is unavailable. Note, however, that $\mathcal C \subset S^\infty(\mathbb S^1)$ and since $\mathcal C$ is a manifold, for each $q \in \mathcal C$, $T_q(\mathcal C) \subset T_q(S^\infty(\mathbb S^1))$. The geometry of $S^\infty(\mathbb S^1)$ is essentially the same as $S^\infty([0,1])$, encountered in the case of PDFs, and the inverse-exponential map at any $q$, $\exp^{-1}_{q}: S^\infty(\mathbb S^1) \to T_{q}(S^\infty(\mathbb S^1))$, takes the same form as seen earlier as long as it is not applied to the antipode $-q$. Our approach hence is to construct a projection $\mathcal C \to T_{\bar q}(\mathcal C)$ that approximates a vector in the (linear) tangent space at the mean shape $[\bar q]$. For a $q \in \mathcal C$, this is achieved in three steps:
\begin{enumerate}
	\item[(i)] register $q$ to $\bar q$ using the shape distance $d_\mathcal S$ by computing $q^*=\bm O^*(q,\gamma^*)$ where $\bm O^*, \gamma^* \in SO(2) \times \Gamma$ optimize $d_\mathcal S([q],[\bar q])$; 
	\item [(ii)] project $q^*$ into $T_{\bar q}(S^\infty(\mathbb S^1))$ by computing 
	the inverse-exponential $\delta q^\circ=\exp_{\bar q}^{-1}(q^*)$;
	\item[(iii)] construct a projection of $\delta q^\circ$ onto $T_{\bar q}(\mathcal C)$ as $\delta q^c=\delta q^\circ-\left[\langle\langle  \delta q^\circ,\phi_1\rangle\rangle \phi_1
	+\langle\langle  \delta q^\circ,\phi_2\rangle\rangle \phi_2\right]$, where $\phi_i: \mathbb S^1 \to \mathbb R^2,\ i\in\{1,2\}$ is the orthonormal basis of the two-dimensional normal subspace $N_{\bar q}(\mathcal C)$ at $\bar{q}$ to $S^\infty(\mathbb S^1)$ in $\mathbb{L}^2\big(\mathbb{S}^1,\mathbb{R}^2)$, since $T_{\bar{q}}(S^\infty(\mathbb S^1))=T_{\bar q}(\mathcal C) \oplus N_{\bar q}(\mathcal C)$. 
\end{enumerate}
We encode steps (i)-(iii) as a projection map $\Pi: \mathcal C \to T_{\bar q}(\mathcal C),\ q \mapsto \Pi(q)=\delta q^c$, which represents the required approximation. In most cases in practice, when the variability amongst the $q_i$ is not large, step (iii) can be avoided. 

In order to obtain the finite-dimensional tangent coordinates, we first compute $\delta q_i^c=\Pi(q_i),\ i\in\{1,\ldots,n\}$. As with PDFs, a data-driven basis for $T_{\bar q}(\mathcal C)$ is obtained by carrying out traditional FPCA with $\delta q_i^c$; in practice, each  $\delta q_i^c$ is sampled using $N$ points and re-shaped into a $2N \times 1$ column vector by stacking the $x$ and $y$ coordinates. Following an SVD of the corresponding sample covariance matrix constructed using the $2N$-dimensional vectors, we compute an $r$-dimensional vector of PC coefficients, $\bm c_i,\ i\in\{1,\ldots,n\}$, following truncation of the eigenbasis to $r$ components.

Starting with a sample, $\beta_1,\ldots,\beta_n$, of closed planar curves, the sequence of operations used to obtain the requisite coordinates is
\begin{equation}
	\label{eg:curve_steps}
	\{\beta_i\} \overset{SRVF}\longrightarrow\{q_i\} \thickspace\overset{\text{mean}}\longrightarrow \thickspace\bar q \thickspace\overset{\Pi(q_i)}\longrightarrow \thickspace\{\delta q^c_i\} \overset {\text{FPCA on }T_{\bar q}(\mathcal C)}\longrightarrow \{\bm c_i \in \mathbb R^r\} \thinspace.
\end{equation}


\section{Tangent functional canonical correlation analysis}
\label{sec:tcca}

In this section, we describe the proposed TFCCA methods. In addition, we also consider canonical variate regression (CVR) \cite{luo2016canonical} for simultaneous estimation of canonical variates and prediction, when a response variable is available. We first briefly discuss the existence of TFCCA. In particular, we elucidate how the situation is distinguished from that of standard FCCA, mainly through the explicit dependence on the point to whose tangent space attention is restricted to.

\subsection{Challenges to establishing existence of FCCA on a tangent space}
\label{sec:existence}
The manifolds $\Psi$ and $\mathcal C$ are submanifolds of Hilbert manifolds $S^\infty([0,1])$ and $S^\infty(\mathbb S^1)$, respectively, where the tangent space at a point is a subspace of a Hilbert space; let $f$ represent a generic point on one of these two spaces. By projecting samples onto the tangent space of a fixed point $f$ on the manifold, we essentially ignore the manifold structure while performing FCCA. Conditions that ensure existence of standard FCCA on Hilbert spaces (see, e.g., \cite{HK}) can be examined in the present setting by explicitly accounting for the presence of $f$.

The point $f$ can be described using two coordinate systems: (i) as the origin in intrinsic local tangent space coordinates of $T_f$, or (ii) as a point in the extrinsic ambient space coordinates which embeds $\Psi$ or $\mathcal C$ into the larger Hilbert space. Local transformations between the two coordinate systems are nonlinear and incompatible with the linear nature of FCCA.

To understand how $f$ influences FCCA, we consider the extrinsic ambient coordinates and identify a subset $A_f$ in the larger Hilbert space whose elements are tangent-like to $f$, and discuss the role of $f$ in adapting existence results of standard FCCA to the present setting. For simplicity, we restrict discussion to PDFs with a non-random $f$, denoted by $\psi$ in our setting (as opposed to an estimated Karcher mean); key arguments carry over to shape curves as well. 


We first briefly review the situation with standard FCCA on the Hilbert space $\mathbb L^2([0,1])$.
The chief obstacle in extending CCA from the multivariate to the functional setting lies in the fact that $\mathbb L^2([0,1])$ is too large and one needs to restrict to a smaller subspace. Specifically, consider two centered random functions $X$ and $Y$, assuming values in two Hilbert spaces $H_1=\mathbb L^2([0,1])$ and $H_2=\mathbb L^2([0,1])$, respectively, with covariance and cross covariance operators $\mathbb K_{lm}, \ l,m\in\{1,2\}$; $\mathbb K_{11}$ and $\mathbb K_{22}$ are assumed to be symmetric, positive definite and Hilbert-Schmidt, whereas $\mathbb K_{12}$ and $\mathbb K_{21}$ are Hilbert-Schmidt.

Let $\{(\lambda_i,\phi_i)\}$ and $\{(\zeta_j,\theta_j)\}$ be eigenvalue-function pairs of $\mathbb K_{11}$ and $\mathbb K_{22}$, respectively, leading to the Karhunen-Lo\'{e}ve representations $X(t)=\sum_i a_i\phi_i(t)$ and $Y(t)=\sum_{i}c_i\theta_i(t),\ t \in [0,1]$, 
where the $a_i$ are uncorrelated with $E(a_i)=0$ and $E(a_i^2)=\lambda_i$, $c_i$ are uncorrelated with $E(c_i)=0$ and $E(c^2_i)=\zeta_i$, and $\{\phi_i\}$ and $\{\theta_i\}$ are orthonormal bases of $H_1$ and $H_2$, respectively. When they exist, the $k$th correlation $\rho_k$ and weight functions $v_k$ and $w_k$ are defined as
\begin{align*}
	\rho_k=\langle\langle v_k,\mathbb K_{12}(w_k)\rangle\rangle =\sup \big\{\langle\langle v,\mathbb K_{12}(w)\rangle\rangle: v,w \in \mathbb L^2[0,1]\big\},
\end{align*}
subject to $\langle\langle v,\mathbb K_{11}(v) \rangle\rangle = \langle\langle w,\mathbb K_{22}(w)\rangle\rangle =1$, and the set of pairs $(\langle\langle v_i, X\rangle\rangle, \langle\langle w_i,Y\rangle\rangle)$ and $(\langle\langle v_j,X\rangle\rangle, \langle\langle w_j,Y\rangle\rangle)$ being uncorrelated for all $ i \neq j$.

A sufficient condition for the existence of $\rho_k$ and weight functions $v_k,w_k$ can be found, for instance, in \cite{HK}. Summarily, assume that $\lambda_i>0$ and $\zeta_j>0$ for every $i,j\in\{1,2,\dots\}$, which ensures invertibility of $\mathbb K_{ll}$ and its square root $\mathbb K_{ll}^{1/2}$ for $l\in\{1,2\}$. The existence of FCCA is characterized by the existence of two correlation operators $\mathbb K^{-1/2}_{11} \mathbb K_{12}\mathbb K^{-1/2}_{22}: \text{range}(\mathbb K^{1/2}_{22}) \to \text{range}(\mathbb K^{1/2}_{11})$ and $\mathbb K^{-1/2}_{22} \mathbb K_{21}\mathbb K^{-1/2}_{11}: \text{range}(\mathbb K^{1/2}_{11}) \to \text{range}(\mathbb K^{1/2}_{22})$, where $\text{range}(\mathbb K^{1/2}_{11})=\Big\{ x \in \mathbb L^2([0,1])\big| \sum_{i} \lambda_i^{-1}\langle x, \phi_i\rangle^2  <\infty\Big\}$
and similarly for $\text{range}(\mathbb K^{1/2}_{22})$. Existence of the correlation operators is guaranteed if the following condition is satisfied: 
\begin{equation}
	\label{condition}
	\sum_{i,j}\frac{E^2(a_ic_j)}{\lambda_i^{2}\zeta_j} <\infty, \quad \sum_{i,j}\frac{E^2(a_ic_j)}{\lambda_i\zeta_j^2} < \infty.
\end{equation}

It is possible to ensure that a condition similar to \eqref{condition} holds for random functions that are tangent-like to two fixed points $\psi_1,\psi_2 \in \Psi$ by further controlling the behavior of $\lambda_i$ and $\zeta_j$ in a manner that depends on the functions $\psi_1$ and $\psi_2$. From Section \ref{sec:frame_pdf}, the tangent space at $\psi \in \Psi$, in local coordinates, is given by $T_\psi(\Psi)=\{\delta\psi \in \mathbb L^2([0,1])| \langle\langle \delta\psi,\psi\rangle\rangle=0\}$. We instead consider $\psi$ in the embedding coordinates and note that the set $A_\psi=\big\{y-\langle\langle y,\psi\rangle\rangle \psi, \ \ y \in \mathbb L^2([0,1])\big\}$
is a proper subset of the Hilbert space within which $\Psi$ is embedded with $\langle \langle y, \psi\rangle \rangle=0$ for every $y \in A_\psi$. The set $A_\psi$ is a subset within the embedding space whose elements behave like tangent vectors to $\psi$. Accordingly, given random functions $X$ and $Y$ considered above, and two fixed $\psi_1,\psi_2 \in \Psi$, consider the random functions $U(t)=X(t)-\langle\langle X,\psi_1 \rangle\rangle \psi_1(t),\   V(t)=Y(t)-\langle\langle Y,\psi_2 \rangle\rangle \psi_2(t)$ that assume values in $A_{\psi_1}$ and $A_{\psi_2}$, respectively.
Consequently, (realizations of) $X,U$ and $\psi_1$ are viewed as elements of $H_1=\mathbb L^2([0,1])$, whereas (realizations of) $Y,V$ and $\psi_2$ are points in $H_2=\mathbb L^2([0,1])$. 
The random functions $U$ and $V$ are centered with covariance operators that depend on $\mathbb K_{lm},\ l,m\in\{1,2\}$ in a nonlinear manner. 

Let $\psi_1(t)=\sum_{i}b_i\phi_i(t)$ and $\psi_2(t)=\sum_{j}d_j\theta_j(t)$ with non-random real coefficients, which satisfy $\sum_{i}b_i^2=\sum_{j}d_j^2=1$. We identify a function in $\mathbb L^2([0,1])$ with a square summable sequence in $l^2$ in the usual manner. Accordingly, define the $l^2$ sequence $\bm a=(a_1,a_2,\ldots)$, and in similar fashion define $\bm b, \bm d$ and $\bm c$; note that $\bm a$ and $\bm c$ are random while $\bm b$ and $\bm d$ are non-random. 
Then, $U(t)=\sum_i \xi_i\phi_i(t),\ V(t)=\sum_j \eta_j \theta_j(t)$, where $\xi_i=a_i- \langle \bm a, \bm b\rangle b_i$ and $\eta_j=c_j-  \langle \bm c, \bm d\rangle d_j$ with $E(\xi_i)=E(\eta_j)=0$ for all $i$ and $j$. Also,
\begin{align*}
	E(\xi_i^2)&=E\left(a_i^2+ \langle \bm a, \bm b\rangle^2b_i^2-2 \langle \bm a, \bm b\rangle a_ib_i \right)
	=\lambda_i(1-b_i^2)^2+b_i^2 \sum_{j \neq i} \lambda_jb_j^2\thickspace.
\end{align*}
In similar fashion, we obtain $E(\eta_j^2)=\zeta_j(1-d_j^2)^2+d_j^2 \sum_{i \neq j} \zeta_id_i^2$. Then,
\begin{align*}
	E(\xi_i\eta_j )&=E\big(a_ic_j-a_id_j \langle \bm c ,\bm d\rangle
	-b_ic_j \langle \bm a ,\bm b\rangle+b_id_j\langle \bm a, \bm b\rangle \langle \bm c ,\bm d\rangle\big)\\
	&=\alpha_{ij}-d_j\langle \bm{\alpha}_i, \bm d\rangle-b_i\langle \bm{\alpha}_j, \bm b\rangle+b_id_i \langle \bm d, \mathbb K_{12}\bm (\bm b) \rangle\thinspace,
\end{align*}
where $\alpha_{ij}=E(a_ic_j)$, $\bm \alpha_i=(\alpha_{i1},\alpha_{i2},\ldots)$, $\bm \alpha_j=(\alpha_{1j},\alpha_{2j},\ldots)$, and through an abuse of notation $\langle \bm d, \mathbb K_{12}\bm (\bm b) \rangle:=E(\langle \langle \bm \psi_1,X \rangle \rangle \langle \langle \bm \psi_2,Y \rangle \rangle )=E(\langle \bm a, \bm b\rangle \langle \bm c ,\bm d\rangle)$. Existence of FCCA on a tangent space then depends on (i) relative signs of $\alpha_{ij}$ and $E(\xi_i\eta_j)$ and (ii) behaviour of $d_j\langle \bm{\alpha}_i, \bm d\rangle-b_i\langle \bm{\alpha}_j, \bm b\rangle+b_id_i \langle \bm d, \mathbb K_{12}\bm (\bm b) \rangle$, which depends on the non-random points $\psi_1$ and $\psi_2$.  

We consider one of the possible scenarios to further explicate on the role of the non-random functions. Suppose $\alpha_{ij}$ and $E(\xi_i\eta_j)$ are both positive. 
If one then assumes that $[d_j\langle \bm{\alpha}_i, \bm d\rangle+b_i\langle \bm{\alpha}_j, \bm b\rangle-b_id_i \langle \bm d, \mathbb K_{12}\bm (\bm b)\rangle]>0$, the numerator of \eqref{condition} can be bounded since $E(\xi_i \eta_j) \leq E(a_ic_j)=\alpha_{ij}$. Since by assumption in standard FCCA, $\lambda_i>0$ and $\zeta_j>0$, we obtain
$E(\xi_i^2) \geq \lambda_i(1-b_i^2)^2$ and $E(\eta_j^2) \geq \zeta_j(1-d_j^2)^2$, and thus
\[
\frac{E^2(\xi_i\eta_j)}{[E^2(\xi^2_i) E(\eta^2_j)]} \leq \frac{E^2(a_ic_j)}{\lambda^2_i\zeta_j}(1-b_i^2)^{-4}(1-d_j^2)^{-2}
\leq M_1\frac{E^2(a_ic_j)}{\lambda^2_i\zeta_j},
\]
and 
\[
\frac{E^2(\xi_i\eta_j)}{[E(\xi^2_i) E^2(\eta^2_j)]} \leq \frac{E^2(a_ic_j)}{\lambda_i\zeta^2_j}(1-b_i^2)^{-2}(1-d_j^2)^{-4}
\leq M_2\frac{E^2(a_ic_j)}{\lambda_i\zeta^2_j},
\]
for some finite constants $M_1$ and $M_2$, since $b_i^2<1$, $d_j^2<1$ and $\sum_i b_i^2=\sum_j d_j^2=1$.
Consequently, a tangent FCCA depending on the nonrandom $\psi_1$ and $\psi_2$ exists when combined with \eqref{condition}. Similar conditions can be formulated when when signs of $\alpha_{ij}$ and $E(\xi_i\eta_j)$ are negative or are different; but the main point is that the essentially linear nature of FCCA implies that the points $\psi_1$ and $\psi_2$ on the nonlinear manifold at which its local linearizations are considered influence the consequent linear correlations between tangent functions in a non-trivial manner. 

A similar conclusion continues to hold, under an appropriate modification, if only the set $A_\psi$ corresponding to a fixed point $\psi$ is used to define both random functions $U$ and $V$ with $\bm b =\bm d$. However, the more realistic situation with $\psi_1$ and $\psi_2$ chosen in a data-driven manner (e.g., Karcher means), or random in the population-level formulation, is more complicated and is deserving of further consideration.
\subsection{Tangent FCCA for PDFs and shape curves}
\label{sec:tcca_pdf}
Once the functional data objects have been projected onto a subspace spanned by an $r$-dimensional PC basis of the tangent space at the sample Karcher mean following the sequence of operations given by \eqref{eq:pdf_steps} or \eqref{eg:curve_steps}
described in Sections \ref{sec:frame_pdf} and \ref{sec:frame_shape}, respectively, traditional multivariate CCA can be carried out on the $\bm c_i \in \mathbb R^r$. We briefly describe how this is carried out in general terms for two paired groups of PDFs or shape curves, or PDFs and shape curves. Given two paired groups of functional data objects $f_{11},\dots,f_{1n}$ and $f_{21},\dots,f_{2n}$, three different options present themselves:
\begin{enumerate}
	\item [(i)] Compute Karcher means $\bar f_1$ and $\bar f_2$ for the two groups, project sample from group 1 into tangent space at $\bar f_1$ and similarly for group 2, and obtain $\bm c_{1i}$ and $\bm c_{2j}$ from groupwise eigenbases on separate tangent spaces. 
	\item [(ii)] Compute a pooled sample Karcher mean $\bar f$ by combining the two samples, project both samples onto the tangent space at $\bar f$, and compute a common eigenbasis to obtain coefficient vectors $\bm c_{1i}$ and $\bm c_{2j}$.
	\item [(iii)] Compute Karcher means $\bar f_1$ and $\bar f_2$ for the two groups, project sample from group 1 into tangent space at $\bar f_1$ and similarly for group 2. Then, parallel transport the tangent vectors from tangent space of, say, $\bar f_1$ onto tangent space of $\bar f_2$, compute a joint eigenbasis, and obtain $\bm c_{1i}$ and $\bm c_{2j}$. More details on parallel transport for the relevant spaces can be found in \cite{srivastava2016functional}. In short, parallel transport preserves the length of tangent vectors and angles between them.
\end{enumerate}
Note that options (ii) and (iii) cannot be used for a paired sample of PDFs and shape curves due to their different representation spaces. We examine the three options on simulated examples in Section \ref{sec:simul} for PDFs and Section 1 in the supplement for shapes. In Section \ref{sec:empirical}, we use option (i) for analysis of real data from multimodal MRIs. Once the vectors $\bm c_i \in \mathbb R^r$ have been obtained for both groups this distinction is vacuous since multivariate CCA is carried out in practice. However, when visualizing canonical variate directions on the space of PDFs or shape curves the choice between options (i)-(iii) assumes relevance. 

The use of a common dimension $r$ for the two groups is not necessary, but we adopt it for simplicity. The dimension $r$ is chosen based on the percentage of variance explained by the corresponding FPCA eigenbasis. Regardless of the option used to obtain the $\bm c_{ki},\ k\in\{1,2\}$, consider matrices $\bm C_1,\ \bm C_2\in\mathbb{R}^{n \times r}$, where each row in these matrices is a vector $\bm c_i$. We assume a common sample size $n$ for both groups, although this can be quite easily relaxed. We find the first pair of canonical weight vectors, $(\bm w_{11},\bm w_{21}) \in \mathbb{R}^r$ for the two groups by solving the following optimization problem:
\begin{align}\label{cca}
	(\bm w^*_{11},\bm w^*_{21}) &= \argmax_{\bm w_{11},\bm w_{21}} \texttt{corr}\big(\bm C_1 \bm w_{11},\bm C_2 \bm w_{21}\big) = \argmax_{\bm w_{11},\bm w_{21}} \frac{\bm w^T_{11} \bm C^T_1 \bm C_2 \bm w_{21}}{\sqrt{\bm w^T_{11} \bm C^T_1 \bm C_1 \bm w_{11}}\sqrt{\bm w^T_{21} \bm C^T_2 \bm C_2 \bm w_{21}}} \thickspace.
\end{align}

\noindent This is the classical formulation of CCA, which finds two vectors that maximize the sample correlation $\rho_1$ between two linear combinations $\bm C_1 \bm w_{11}$ and $\bm C_2 \bm w_{21}$, i.e., the canonical variates. Once we obtain the first pair $(\bm w^*_{11},\bm w^*_{21})$, we successively find subsidiary canonical weight vectors. Thus, the $j$th pair $(\bm w_{1j},\bm w_{2j})$ is found to maximize the $j$th sample correlation,
\begin{align*}
	&\rho_j = \max_{\bm w_{1j},\bm w_{2j}} \texttt{corr} \big( \bm C_1 \bm w_{1j}, \bm C_2 \bm w_{2j} \big) ~~~ \text{subject to the constraints}  \\
	\texttt{corr} & \big( \bm C_1 \bm w_{1j}, \bm C_1 \bm w_{1l} \big) = 0,~\texttt{corr} \big( \bm C_2 \bm w_{2j}, \bm C_2 \bm w_{2l} \big) = 0,~\texttt{corr} \big( \bm C_1 \bm w_{1j}, \bm C_2 \bm w_{2l} \big) = 0,
\end{align*}

\noindent for $j \ne l$ and $j, l \in\{1,\ldots,r\}$. If the dimensions of $\bm C_1$ and $\bm C_2$ are different, we can compute sample correlations, and associated canonical variates, up to the minimum rank of $\bm C_1$ and $\bm C_2$. 

A key element of our approach, especially in the context of the medical imaging application described in Section \ref{sec:motapp}, is the ability to visualize the estimated canonical variates directly on the space of PDFs or shape curves. We describe how this is done when  option (ii) with a pooled sample Karcher mean $\bar f$ is used, but the description is valid across the three options with obvious modifications. 
To visualize the $j$th canonical variate direction for group $k$, we first construct the corresponding tangent vector $v_{kj}\in T_{\bar f}(M)$ ($M$ here corresponds to either $\Psi$ or $\mathcal{C}$) using $v_{kj}(t)=\sum_{i=1}^r e_{n,i}(t) w^*_{kji}$,
where $\{e_{n,i}\},\ i\in\{1,\ldots,r\}$, are the first $r$ estimated eigenfunctions of the sample covariance operators described in Section \ref{sec:frame}.
When considering PDFs, the estimated canonical variate direction is given by
\begin{equation}\label{tcca1}
	p_{kj}(t;\epsilon) = \left[\exp_{\bar{\psi}}\big(\epsilon v_{kj} \big)\right]^2(t),
\end{equation}
where $\epsilon$ is a parameter that scales the length of the geodesic from the mean $\bar \psi$ to $\exp_{\bar \psi}(v_{kj})$; the squaring operation transforms back to the space of densities $\mathcal P$ from $\Psi$. The exponential map is easy to compute since it is available in closed-form as shown in Section \ref{sec:frame}. For shape curves with mean $\bar{q}\in\mathcal{C}$, the projection of $v_{kj}$ from $T_{\bar q}(\mathcal C)$ to $\mathcal C$ is done numerically; we refer to \cite{srivastava2016functional} for details. The parameter $\epsilon$ is typically selected as a set of integers, e.g., $-2,-1,0,1,2$. For each direction, in order to visualize it, we move a certain amount in the ``positive'' and ``negative'' directions from the mean as specified by $v_{kj}$.

\subsection{Regression using canonical variates}
\label{sec:cvr}

We describe a regression framework that extends the proposed TFCCA, commonly referred to as canonical variate regression (CVR). The CVR approach was proposed by Luo et al. \cite{luo2016canonical}, and was motivated by the work of Gross and Tibshirani \cite{gross2015collaborative}. It combines two cost functions to jointly estimate canonical weight vectors as well as regression parameters. The first cost function favors canonical weight vectors that maximize the correlation between resulting canonical variates. The second cost function, resulting from an appropriate log-likelihood, favors canonical weight vectors that minimize the regression residuals. 

Specifically, let $\bm C_1,\ \bm C_2 \in \mathbb{R}^{n \times r}$ correspond to FPC coefficient matrices for two groups. Also, let $\bm Y\in\mathbb{R}^n$ denote the response vector. In CVR, we aim to solve the following optimization problem: 
\begin{align}\label{cvr}
	&\argmin_{\bm W_{1},\bm W_{2}, \alpha,\bm \beta} \Big\{  \eta \big| \bm C_1 \bm W_{1} - \bm C_2 \bm W_{2} \big|_F^2 ~ + ~ (1-\eta) \sum^2_{k=1} l(\alpha,\bm \beta; \bm Y,\bm C_k\bm W_k)\Big\}
\end{align}
subject to $\bm W^T_1\bm C^T_1\bm C_1\bm W_1=\bm W^T_2\bm C^T_2\bm C_2\bm W_2=\bm I_d$, where $\bm W_1,\bm W_2\in \mathbb{R}^{r\times d}$ ($d\leq r$ is a pre-specified number of desired canonical variates) are the canonical weight matrices, $|\cdot|_F$ is the Frobenious norm, $l(\alpha,\bm \beta; \bm Y,\bm C_k\bm W_k)=\big| \bm Y - (\alpha\mathbf{1}_{n} + \bm C_k \bm W_{k}\bm \beta) \big|_{F}^2$ is the linear regression cost function, and $\alpha\in\mathbb{R}$ and $\bm \beta\in\mathbb{R}^d$ are the regression model parameters. The first term in \eqref{cvr} is related to the correlation between the canonical variates $\bm C_1 \bm W_{1}$ and $\bm C_2 \bm W_{2}$, while the second term captures prediction error. If $\eta=1$, the CVR model reduces to classical CCA. When $\eta=0$, the CVR model becomes a linear regression model with canonical variates as predictors. The parameter $\eta$ is generally tuned via cross-validation based on some criterion such as the mean squared error (MSE). The joint optimization problem in \eqref{cvr} is solved via the alternating direction method of multipliers (ADMM) algorithm \cite{boyd2011distributed}.


\section{Simulation study for PDFs}
\label{sec:simul}

In this section, we evaluate the proposed TFCCA procedure for PDFs. A similar study for shapes is included in Section 1 in the supplement. We design three datasets of PDFs to assess performance of the proposed TFCCA approach. When computing the low-dimensional, Euclidean representation of PDFs via their FPC coefficients, we compare the performance of options (i) (separate tangent spaces) and (ii) (common tangent space at the pooled Karcher mean). The main goal of the presented simulation is to assess the accuracy of the estimated canonical correlations and variates. Simulating ``ground truth" canonical correlations (variates) between PDFs that correspond to our method is tricky since we are not given a tangent space and projected vectors a priori. We hence simulate in the following manner, depending on the option of common or separate tangent spaces from Section \ref{sec:tcca_pdf}:
\begin{enumerate}
	\item Simulate two groups of 100 random vectors $\bm x_{ik}=(x_{1ik},\ldots,x_{rik})^T,\ i\in\{1,\ldots,100\},\ k\in\{1,2\}$ from an $r$-dimensional Gaussian distribution with a pre-specified mean and covariance. Implement multivariate CCA on these to obtain ``ground truth" canonical correlations, $\bm \rho$, and canonical variates, $\bm C \bm W$. 
	\item Simulate two groups of PDFs as specified in the next paragraph. 
	\item Obtain eigenbasis on tangent space at mean PDF. 
	\begin{enumerate}
		\item [(a)] For option (i) of using separate tangent spaces, compute different Karcher mean PDFs and project the PDFs onto the different tangent spaces; carry out separate FPCA to get $r$ eigenfunctions on each tangent space;
		\item [(b)] For option (ii) of using a common tangent space, pool the samples, compute a single Karcher mean PDF, project both samples of PDFs onto tangent space at this mean, and compute an $r$-dimensional joint eigenbasis for the two groups.
	\end{enumerate}
	\item Simulate two new groups of PDFs using the same $\bm x_{ik}$ as in Step 1 and eigenbases from Step 3. For group $k$, simulate the $i$th PDF as $p_{ik}=\exp_{\bar \psi_k}(\sum_{j=1}^r x_{jik}e_{j})^2$. Here, 
	$\bar \psi_k$ is the mean for the $k$th group and $e_{j}$ are the eigenfunctions for group $k$.
	\item Treating the two new groups of PDFs $\{p_{ik}\}$ as data, estimate mean(s), project to tangent space(s), carry out FPCA, and obtain coefficient matrices $\bm C_1$ and $\bm C_2$, in a manner that is compatible with choosing separate or common tangent spaces. 
	\item Carry out multivariate CCA on $\bm C_i,\ i\in\{1,2\}$ and obtain vector of estimated canonical correlations, $\hat{\bm \rho}_c$ and $\hat{\bm \rho}_s$, and canonical variates, $\hat{\bm C}_c \hat{\bm W}_c$ and $\hat{\bm C}_s \hat{\bm W}_s$, depending on whether a common tangent space is, or separate tangent spaces are, used, respectively. 
\end{enumerate}

\begin{figure}[!t]
	\begin{center}
		\begin{tabular}{|c|c|c|}
			\hline
			Group 1 & Group 2 & Group 3 \\
			\hline
			\begin{minipage}{.25\textwidth}
				\includegraphics[width=\linewidth]{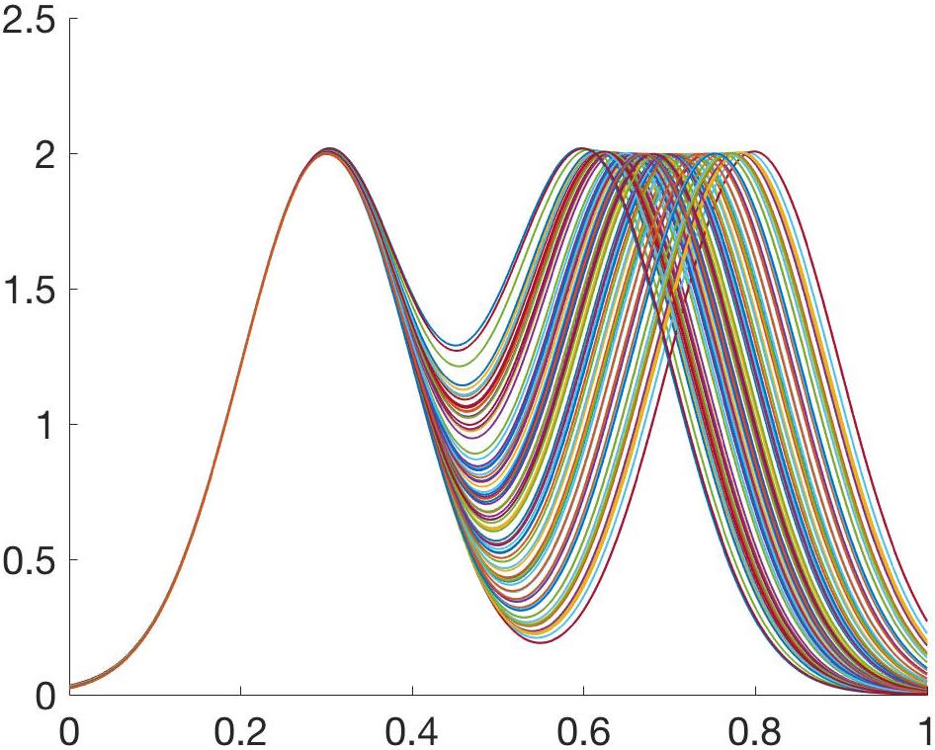}
			\end{minipage}
			&
			\begin{minipage}{.25\textwidth}
				\includegraphics[width=\linewidth]{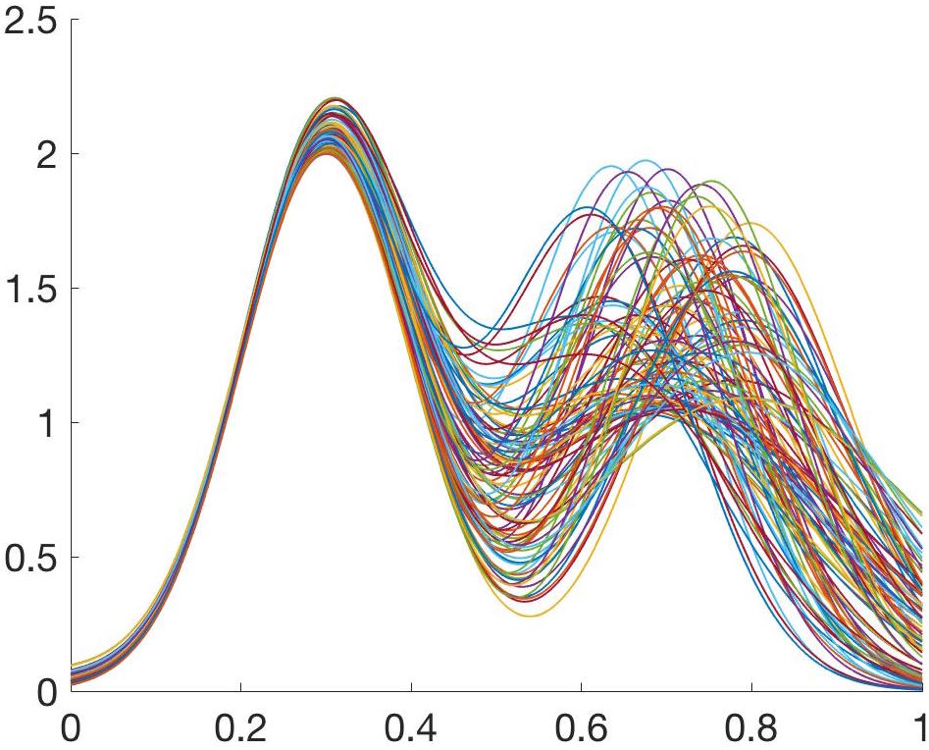}
			\end{minipage}
			&
			\begin{minipage}{.25\textwidth}
				\includegraphics[width=\linewidth]{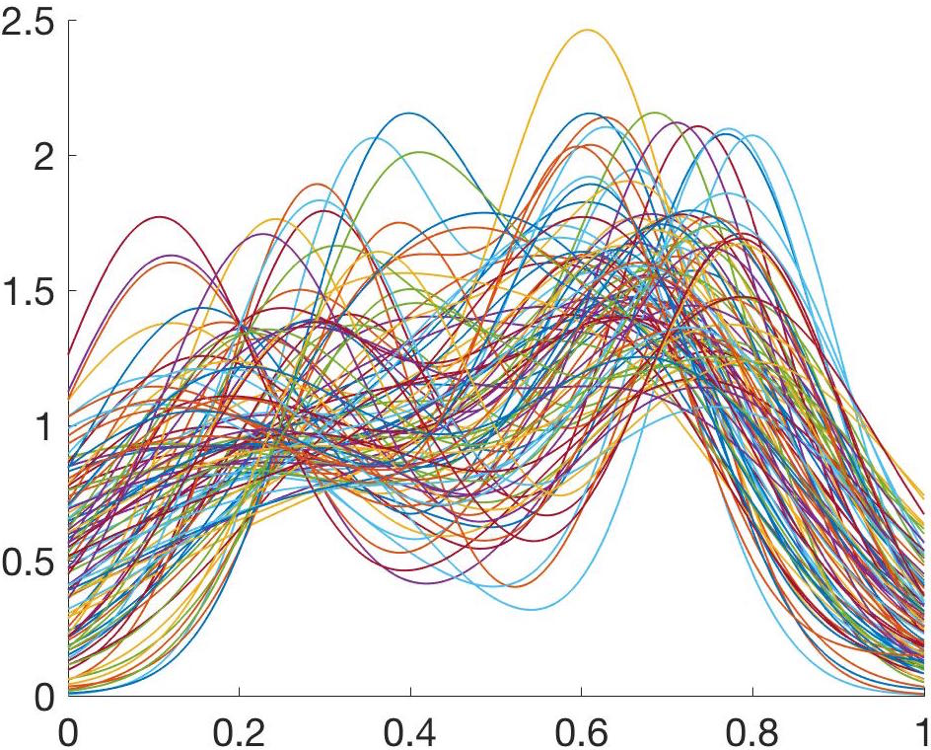}
			\end{minipage}
			\\
			\hline
		\end{tabular}
	\end{center}
	\caption{100 simulated PDFs from each of three different groups defined in Section \ref{sec:simul}.}
	\label{fig:simpdf}
\end{figure}

We generate three different groups of 100 random PDFs using a mixture of two Gaussian distributions truncated to $[0, 1]$ as shown in Fig. \ref{fig:simpdf}. Each PDF is represented using $N=1000$ equally spaced points. The Gaussian mixture with various choices of mean and variance for each group is computed as follows, where $\phi(\mu,\sigma)$ denotes a Gaussian PDF with mean $\mu$ and standard deviation $\sigma$, $0.5~\phi(\mu_1,\sigma_1) ~~ + ~~ 0.5~\phi(\mu_2,\sigma_2)$, with parameters for each group given by: Group 1: $\mu_1 = 0.3,~\mu_2 \sim \text{Unif}(0.6,0.8)$, $\sigma_1 = \sigma_2 = 0.1$; Group 2: $\mu_1 = 0.3,~\mu_2 \sim \text{Unif}(0.6,0.8)$, $\sigma_1 = 0.1,~ \sigma_2 \sim \text{Unif}(0.1,0.2)$; Group 3: $\mu_1 \sim \text{Unif}(0.1,0.4),~ \mu_2 \sim \text{Unif}(0.6,0.8)$, $\sigma_1 \sim \text{Unif}(0.1,0.3),~ \sigma_2 \sim \text{Unif}(0.1,0.2)$.


\begin{table}[!t]
	\caption{Comparison of ``ground truth'' canonical correlations and re-estimated canonical correlations in separate tangent spaces ($\hat{\bm \rho}_s$) and a common tangent space ($\hat{\bm \rho}_c$). We use $r=2$, $r=3$ and $r=4$ for Groups 1 \& 2, 1 \& 3 and 2 \& 3, respectively. The groups are defined in Section \ref{sec:simul}.}
	\label{tab:simpdf}
	\begin{center}
		\begin{footnotesize}
			\makebox[\textwidth]{\begin{tabular}{cccc}
					\hline
					Setting & Groups 1 \& 2 & Groups 1 \& 3 & Groups 2 \& 3 \\
					$\bm \rho$ & $(0.71,0.27)$ & $(0.63,0.26,0.12)$ & $(0.82,0.13,0.11,0.03)$ \\
					$|\bm \rho-\hat{\bm \rho}_s|$ & $(0.11,0.31)\times10^{-5}$ & $(0.10,0.03,0.05)\times10^{-3}$ & $(0.04,0.01,0.10,0.08)\times10^{-3}$ \\
					$|\bm \rho-\hat{\bm \rho}_c|$ & $(0.42,0.44)\times10^{-4}$ & $(0.01,0.02,0.07)\times10^{-2}$ & $(0.01,0.05,0.01,0.01)\times10^{-2}$ \\
					\hline
			\end{tabular}}
		\end{footnotesize}
	\end{center}
\end{table}

Table \ref{tab:simpdf} compares the estimates $\hat{\bm \rho}_c$ and $\hat{\bm \rho}_s$ to the ``ground truth'' $\bm \rho$ based on different selections of the two groups; for Groups 1 and 2 we use $r=2$, for Groups 1 and 3 we use $r=3$, and for Groups 2 and 3 we use $r=4$. For Groups 1 and 2, two FPCs explain over 98$\%$ of the total variation while for Group 3, three FPCs explain over $93\%$ of the total variation. It is evident that differences between $\bm \rho$ and $\hat{\bm \rho}_c$ or $\hat{\bm \rho}_s$ are very small for all scenarios. Table \ref{tab:simpdf2} further compares the estimates $\hat{\bm C}_c \hat{\bm W}_c$ and $\hat{\bm C}_s \hat{\bm W}_s$ of the canonical variates in each group to the ``ground truth'' $\bm C \bm W$ based on different selections of the two groups as in Table \ref{tab:simpdf}. The comparison is made using the standard Euclidean norm after centering. Again, it is evident that differences between $\bm C \bm W$ and $\hat{\bm C}_c \hat{\bm W}_c$ or $\hat{\bm C}_s \hat{\bm W}_s$ are very small in all cases. These results indicate that the recovery of the canonical correlations and canonical variates is very accurate, and both options (i) and (ii) for estimating the FPC coefficient matrices work well. We remark that it is not surprising that the differences between $\bm \rho$ and $\hat{\bm \rho}_c$ ($\bm C \bm W$ and $\hat{\bm C}_c \hat{\bm W}_c$) are larger than the differences between $\bm \rho$ and $\hat{\bm \rho}_s$ ($\bm C \bm W$ and $\hat{\bm C}_s \hat{\bm W}_s$), since we constructed PDFs for each group using their individual truncated eigenbases (in separate tangent spaces). 


\begin{table}[!t]
	\caption{Comparison of ``ground truth'' canonical variates and re-estimated canonical variates in separate tangent spaces ($\hat{\bm C}_s \hat{\bm W}_s$) and a common tangent space ($\hat{\bm C}_c \hat{\bm W}_c$); the additional subscripts correspond to the two different groups. We use $r=2$, $r=3$ and $r=4$ for Groups 1 \& 2, 1 \& 3 and 2 \& 3, respectively. The groups are defined in Section \ref{sec:simul}.}
	\label{tab:simpdf2}
	\begin{center}
		\begin{footnotesize}
			\makebox[\textwidth]{\begin{tabular}{cccc}
					\hline
					Setting & Groups 1 \& 2 & Groups 1 \& 3 & Groups 2 \& 3 \\
					$\|\bm C_1 \bm W_1 -\hat{\bm C}_{1s} \hat{\bm W}_{1s}\|$ & $(0.17,0.16)\times10^{-4}$ & $(0.43,0.53,0.34)\times10^{-4}$ & $(0.23,0.69,0.31,0.81)\times10^{-4}$ \\
					$\|\bm C_2 \bm W_2 -\hat{\bm C}_{2s} \hat{\bm W}_{2s}\|$ & $(0.09,0.14)\times10^{-4}$ & $(0.38,0.62,0.43)\times10^{-4}$ & $(0.51,0.70,0.68,0.80)\times10^{-4}$ \\
					$\|\bm C_1 \bm W_1 -\hat{\bm C}_{1c} \hat{\bm W}_{1c}\|$ & $(0.94,0.57)\times10^{-3}$ & $(0.31,0.42,0.26)\times10^{-2}$ & $(0.15,0.34,0.21,0.14)\times10^{-2}$ \\
					$\|\bm C_2 \bm W_2 -\hat{\bm C}_{2c} \hat{\bm W}_{2c}\|$ & $(0.42,0.46)\times10^{-3}$ & $(0.37,0.23,0.10)\times10^{-2}$ & $(0.29,0.36,0.14,0.17)\times10^{-2}$ \\
					\hline
			\end{tabular}}
		\end{footnotesize}
	\end{center}
\end{table}

\begin{figure}[!t]
	\begin{center}
		\begin{tabular}{|c|c|}
			\hline
			Group 1 & Group 2 \\
			\hline
			\begin{minipage}{.47\textwidth}
				\includegraphics[width=\linewidth]{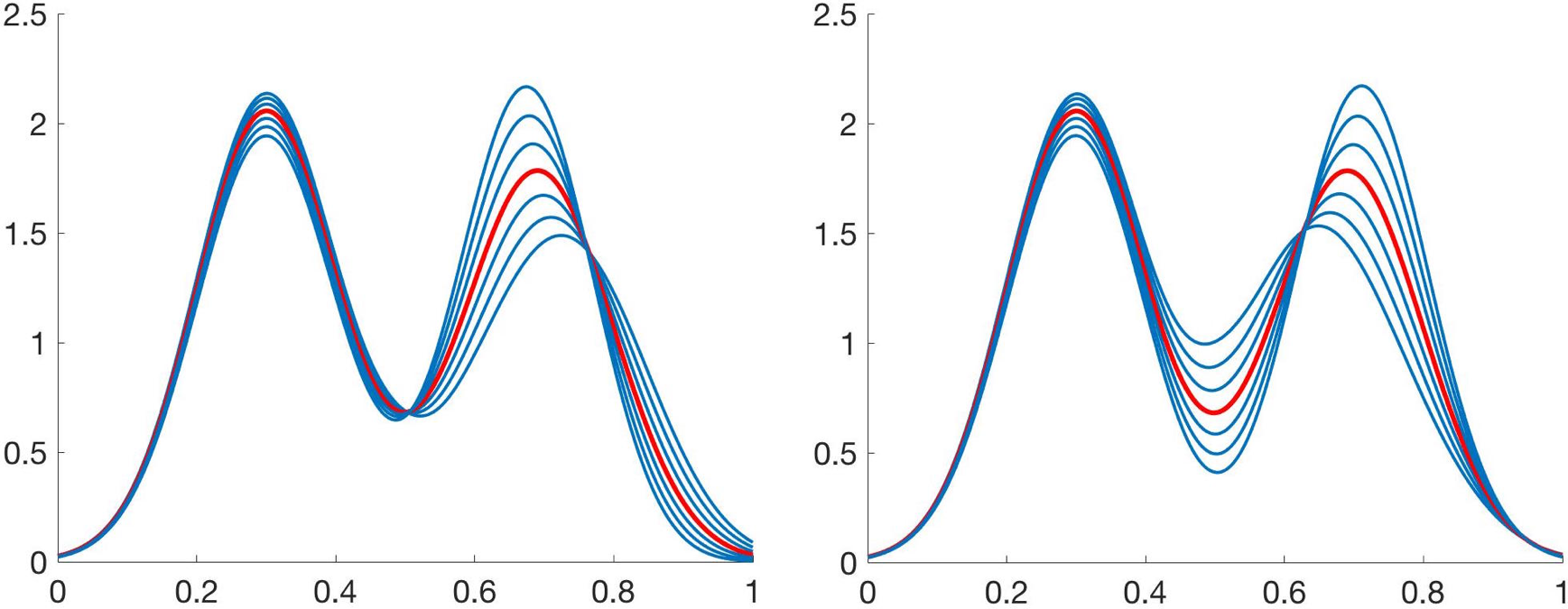}
			\end{minipage}
			&
			\begin{minipage}{.47\textwidth}
				\includegraphics[width=\linewidth]{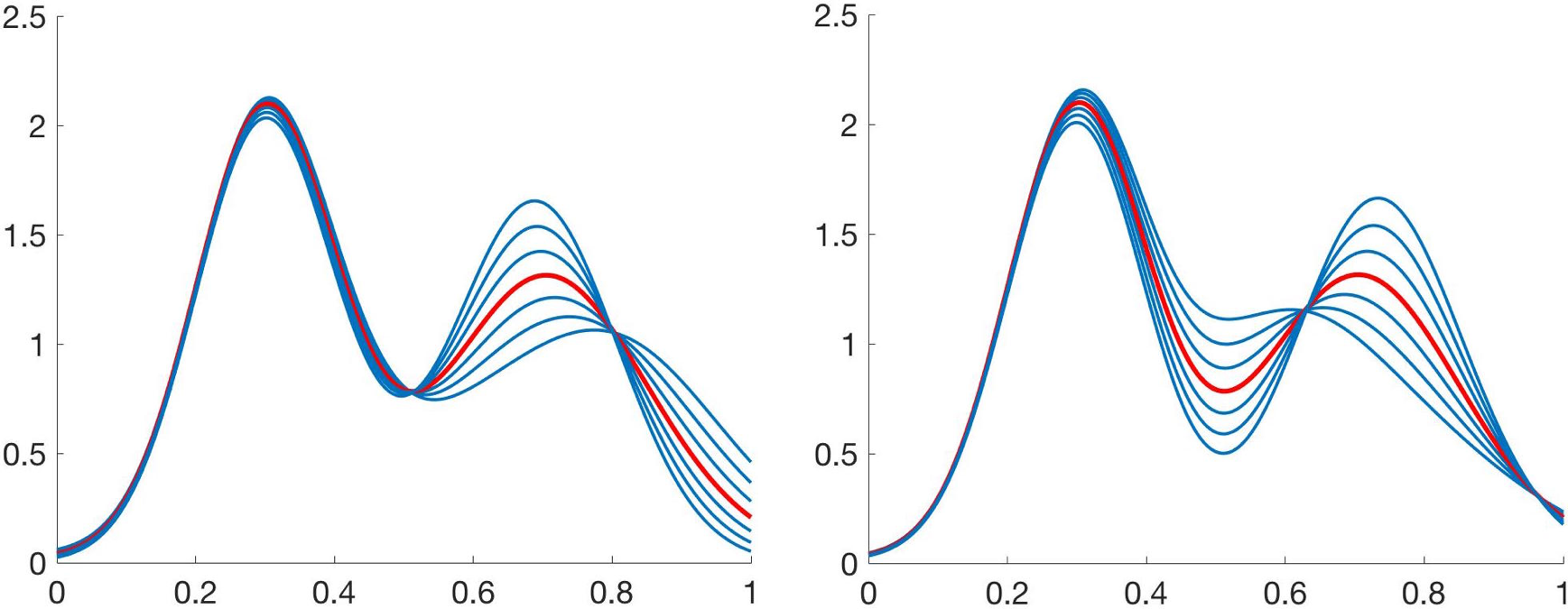}
			\end{minipage}
			\\
			\hline
		\end{tabular}
	\end{center}
	\caption{Visualization of the first two canonical variate directions (left to right) for Groups 1 and 2 estimated on separate tangent spaces. Karcher mean PDFs are in red. For each direction, the six blue curves correspond to PDFs moving along a geodesic from the mean at values $\epsilon\in\{-3,-2,-1,0,1,2,3\}$ as given in \eqref{tcca1}.}
	\label{fig:ccpdf}
\end{figure}

Fig. \ref{fig:ccpdf} presents the first two canonical variate directions for Groups 1 and 2 when estimation is carried out in separate tangent spaces; the two groups are most correlated along these two directions. For each group, the first direction is displayed in the left panel and the second direction is displayed in the right panel. These directions are explored and visualized using a discretized path of PDFs, as defined in \eqref{tcca1}, by setting $\epsilon\in\{-3,-2,-1,0,1,2,3\}$. Although the canonical variate directions appear different for the two groups due to different mean PDFs (red) and FPC eigenbases, the observed patterns are very similar. For both groups, the first direction captures variability in the second peak while the second direction captures variation in the valley and the second peak. In addition, the PDFs along the first direction move from a high second peak to a low second peak, while the PDFs along the second direction move from a high valley with a low second peak to a low valley with a high second peak.


\section{Textural and shape associations of GBM tumors in multimodal MRI\label{sec:empirical}}

Next, we apply the proposed TFCCA approach to study associations between voxel value PDFs, representing the texture of GBM tumors in MRI images, and GBM tumor shapes. Throughout this section, we utilize FPC coefficient-based Euclidean representations of these objects, estimated in separate tangent spaces. We begin with a brief description of the data.

The data used in this work consists of MRIs of GBM brain tumors from 58 patients who consented under The Cancer Genome Atlas protocols (\url{http://cancergenome.nih.gov/}). The GBM data consists of two imaging modalities: pre-surgical T1-weighted post contrast (T1) and T2-weighted fluid-attenuated inversion recovery (FLAIR) (T2) magnetic resonance sequences from The Cancer Imaging Archive (\url{http://www.cancerimagingarchive.net/}). The survival times for each subject were obtained from cBioPortal (\url{http://www.cbioportal.org/}); none of the subjects in our study had censored observations.

The voxel value PDF of a brain tumor is estimated from a histogram of pixel intensities inside a tumor mask. The image pre-processing steps are provided in Fig. 2 of Saha et al. \cite{saha2016demarcate}. For shape analysis, we extracted a closed outer contour of the GBM tumor in the axial MRI slice with the largest tumor area. The tumor rarely possesses landmark features, and thus, the entire contour is needed for analysis. The variances of voxel value PDFs in T1 and T2 are 0.107 and 0.124, respectively. The variances of shape curves of GBM tumors in T1 and T2 are 0.060 and 0.104, respectively. The small variances justify our use of local linearizations of the data in tangent spaces. 

\subsection{TFCCA for PDFs estimated from T1 and T2 modalities}
\label{sec:empirical_pdf}

Given two paired groups of voxel value PDFs corresponding to the T1 and T2 modalities, we estimate the canonical correlations and canonical variates. This assesses the extent to which texture features of GBM tumors are shared by the two MRI modalities. 

\begin{figure}[!t]
	\begin{center}
		\begin{tabular}{|c|}
			\hline
			T1\\
			\hline
			\begin{minipage}{.75\textwidth}
				\includegraphics[width=\linewidth]{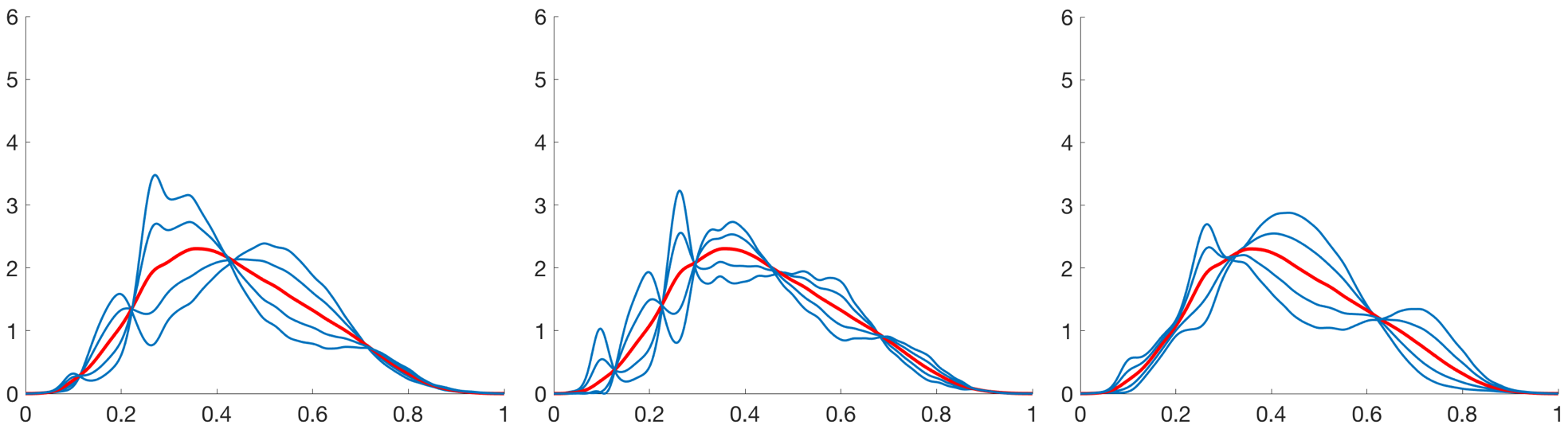}
			\end{minipage}
			\\
			\hline
			T2\\
			\hline
			\begin{minipage}{.75\textwidth}
				\includegraphics[width=\linewidth]{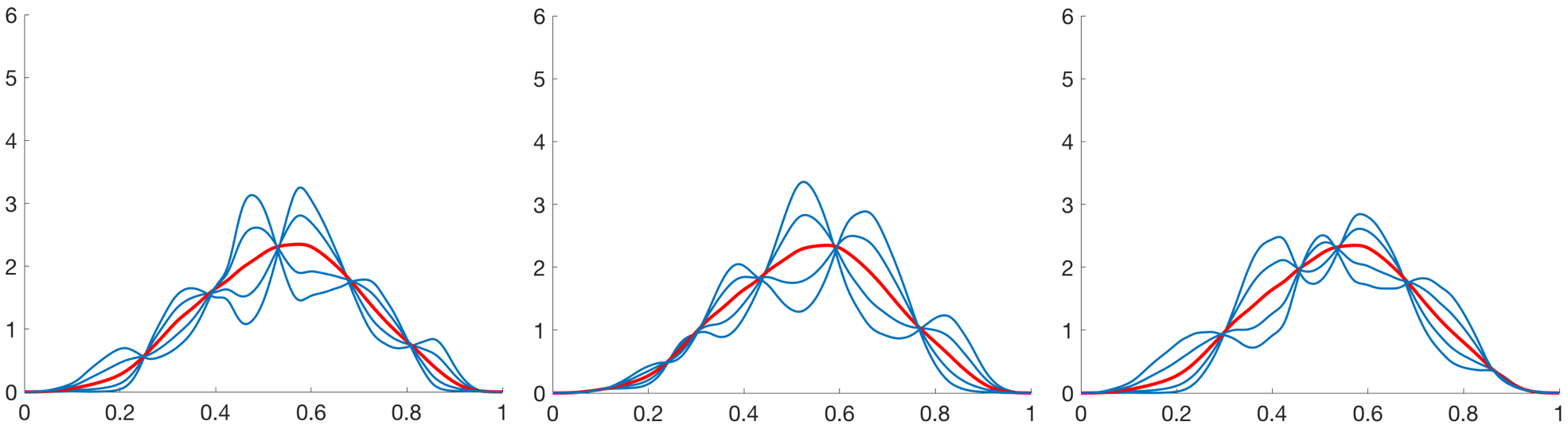}
			\end{minipage}
			\\
			\hline
		\end{tabular}
	\end{center}
	\caption{First three canonical variate directions (left to right) for voxel value PDFs of GBM tumors' appearance in the T1 and T2 modalities. For each direction, the four blue curves correspond to PDFs moving along a geodesic from the mean at values $\epsilon\in\{-2,-1,0,1,2\}$ as given in \eqref{tcca1}.}
	\label{fig:ccs}
\end{figure}

For each modality, the first five FPCs explain over $95\%$ of the total variance. Thus, we construct FPC coefficient matrices $\bm C_1,\ \bm C_2\in \mathbb{R}^{58\times 5}$ using option (i) in Section \ref{sec:tcca_pdf}. We then estimate the leading canonical correlations to be $(0.5900,~0.4747,~0.2405,~0.2260,~0.1155)$. The first two canonical correlations indicate a moderate association between the corresponding canonical variates. Thus, the textural appearance of GBM tumors in the two modalities shares some common characteristics, as captured by the voxel value PDFs. We display the first three canonical variate directions in Fig. \ref{fig:ccs}, from left to right, with $\epsilon\in\{-2,-1,0,1,2\}$; the mean PDF for each modality is shown in red. Recall that these directions represent the linear combinations of FPC directions that maximize their correlations. The first canonical variate direction for the T1 modality appears to capture global shifts in the density of low voxel values to high voxel values; the corresponding canonical variate direction for the T2 modality also captures similar shifts, but they are more local. The second canonical variate direction for the T1 modality mostly captures variation in the density of low voxel values, while for the T2 modality it captures similar variation as the first canonical variate direction. This indicates that the textural variation of tumors tends to have similar characteristics in the two modalities.

\subsection{TFCCA for tumor shapes estimated from T1 and T2 modalities}
\label{sec:empirical_shape}

\begin{figure}[!t]
	\begin{center}
		\begin{tabular}{|c|c|}
			\hline
			T1&T2\\
			\hline
			\begin{minipage}{.45\textwidth}
				\includegraphics[width=\linewidth]{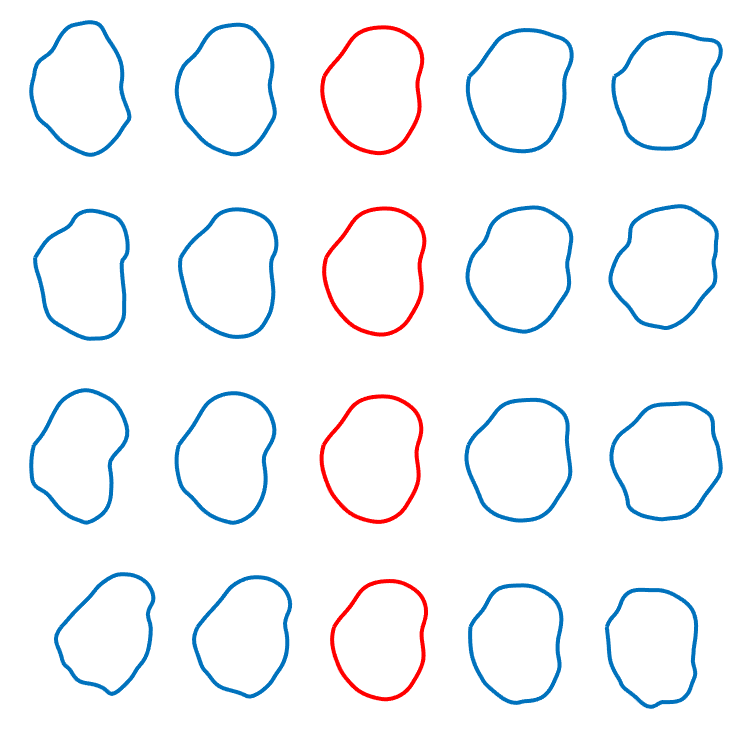}
			\end{minipage}
			&   \begin{minipage}{.45\textwidth}
				\includegraphics[width=\linewidth]{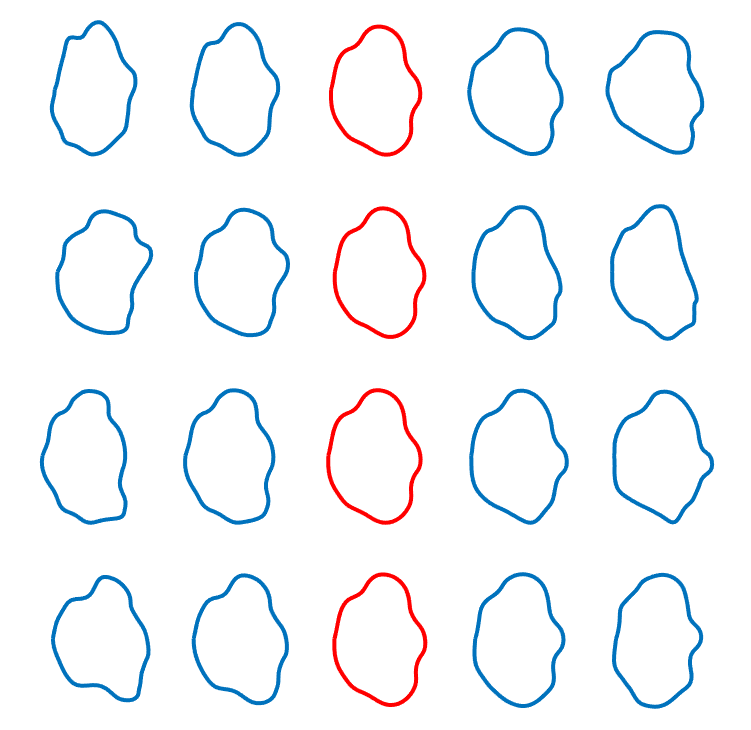}
			\end{minipage}
			\\
			\hline
		\end{tabular}
	\end{center}
	\caption{First four canonical variate directions (top to bottom) for GBM tumor shapes in the T1 and T2 modalities. For each direction, the four blue curves correspond to shapes moving along a geodesic from the mean at values $\epsilon\in\{-2,-1,0,1,2\}$.}
	\label{fig:pccc2}
\end{figure}

Next, we investigate associations between shapes of GBM tumor contours extracted from the T1 and T2 MRI modalities. We first construct FPC coefficient matrices $\bm C_1 \in \mathbb{R}^{58\times 13},\ \bm C_2 \in \mathbb{R}^{58\times 15}$ for the T1 and T2 GBM tumor shapes, respectively, using option (i) in Section \ref{sec:tcca_pdf}. The first 13 FPCs and 15 FPCs for T1 and T2, respectively, capture at least $80\%$ of the total shape variation for each modality. We apply the proposed TFCCA approach to estimate the canonical correlations and corresponding canonical variate directions. The leading canonical correlations are $(0.8284,~0.7916,~0.7522,~0.7249)$; the subsequent canonical correlations are all smaller than 0.7. It appears that the associations between tumor shapes are stronger than the associations between the corresponding voxel value PDFs across the T1 and T2 modalities. Fig. \ref{fig:pccc2} displays the first four canonical variate directions for each modality. The large canonical correlations indicate that the shapes of tumors captured by the two modalities share many common features. To assess GBM tumor severity, radiologists are often interested in protrusions of the tumor into neighboring tissues. It appears that the canonical variate directions estimated using the proposed TFCCA approach capture different types of such protrusions.

Results of TFCCA applied to paired PDFs and shapes from the same MRI modality are presented in Section 2 in the supplement.



\subsection{Canonical variate regression for survival prediction in GBM}
\label{sec:empirical_cvr}

In this section, we use canonical variate regression for simultaneous estimation of canonical variates and survival prediction in the context of GBM. The estimation procedure is carried out separately for voxel value PDFs and shapes.

We begin by representing the T1 and T2 voxel value PDFs using ten FPC coefficients in separate tangent spaces. The corresponding FPCs explain $99.70\%$ and $99.65\%$ of the total variation, respectively. The resulting FPC coefficient matrices, $\bm C_1,\ \bm C_2\in\mathbb{R}^{58\times 10}$ can then be used to estimate canonical variate weights $\bm W_1,\ \bm W_2$ as well as the model parameters $\alpha,\ \bm \beta$ in the CVR model specified in \eqref{cvr}. As a response, we use the natural logarithm of each subject's survival time in months (log-survival). We first randomly split the data into two sets. The first set, which contains $80\%$ of the data, is used for estimation. We repeat estimation for $\eta\in\{0,0.1,0.2,\dots,1\}$. Then, based on the estimates for each value of $\eta$, we compute the mean squared error (MSE) between true log-survival and predicted log-survival using the left-out $20\%$ of the data. We finally retain the estimates corresponding to the value of $\eta$ that generated the minimum MSE. Furthermore, we use the canonical weights estimated using the CVR model to compute canonical variates for the left-out data, and use them as predictors in a Cox proportional hazards model to directly model survival. We assess predictive performance of this model via the commonly used concordance index (C-index) \cite{harrell1982evaluating}. We repeat this procedure 100 times using different $80\%$-$20\%$ splits of the data. The corresponding average MSEs and C-indices (with standard deviations in parentheses) for the different numbers of canonical variates we considered are reported in Table \ref{tab:cvr} (top). We compare this approach to a PC regression method, where ten FPC coefficients from each modality are included as predictors in the regression model. First, our approach always outperforms the PC regression method, both in terms of the average MSE and the average C-index. It appears that the lowest average MSE is attained when only one canonical variate is used for prediction via the CVR model. On the other hand, the average C-index is largest when five canonical variates are used as predictors in the Cox proportional hazards model.

We repeat the same exact procedure for tumor shapes from the T1 and T2 modalities. In this case, we use 20 FPC coefficients to compute the Euclidean coordinates for shapes in each modality; for the T1 modality, the 20 FPCs explain over $92\%$ of the total shape variation, while for the T2 modality, 20 FPCs explain $89\%$ of the total variance. The resulting average MSEs and average C-indices (with standard deviations in parentheses), for different numbers of estimated canonical variates, are reported in Table \ref{tab:cvr} (bottom). In this case, based on the average MSEs, prediction using CVR significantly outperforms prediction using a PC regression model with 20 FPC coefficients per modality. Similarly to the case of voxel value PDFs, the minimum average MSE is achieved when one canonical variate is used. The average C-indices are also larger when canonical variates are used as predictors. However, unlike in the case of voxel value PDFs, the best predictive performance is achieved when one or three canonical variates are used as predictors.

In many cases, the optimal value of the parameter $\eta$ corresponding to the minimum MSE is 0 or 0.1. This places a large weight on the second term in \eqref{cvr}, which tries to minimize prediction error. Thus, the corresponding estimated canonical correlations are generally smaller than those reported in Sections \ref{sec:empirical_pdf} and \ref{sec:empirical_shape}; for a single simulation example with the optimal $\eta=0.1$, they are $(0.5879,~0.4694,~0.2238,~0.2106,~0.0908)$ for PDFs and $(0.8218,~0.7860,~0.7469,~0.7152)$ for shapes. The corresponding canonical variate directions for PDFs and shapes, for each modality, are shown in Figs. 4 and 5 in Section 3 in the supplement. The CVR-based canonical variate directions are slightly different from those displayed in Figs. \ref{fig:ccs} and \ref{fig:pccc2} as they place more weight on predicting log-survival than maximizing correlation. 

\begin{table}[!t]
	\caption{Comparison of predictive performance of CVR (average MSE) and Cox proportional hazards (average C-index) models to a PC regression model based on PDFs (top) and shapes (bottom). Standard deviations are in parentheses.}
	\label{tab:cvr}
	\begin{center}
		\begin{footnotesize}
			\makebox[\textwidth]{\begin{tabular}{c|c|ccccc}
					\hline
					Model   &  10 PCs & 1 CV & 3 CVs & 5 CVs & 7 CVs \\
					MSE     & $0.854~(0.22)$ & $0.596~(0.18)$ & $0.628~(0.20)$ & $0.666~(0.19)$ & $0.733~(0.20)$\\
					C-index & $0.548~(0.08)$ & $0.550~(0.10)$ & $0.554~(0.10)$ & $0.565~(0.10)$ & $0.562~(0.10)$\\
					Model   &  20 PCs & 1 CV & 3 CVs & 5 CVs & 7 CVs \\
					MSE     & $7.633~(2.81)$ & $1.415~(0.53)$ & $1.474~(0.53)$ & $1.564~(0.55)$ & $1.747~(0.64)$\\
					C-index & $0.475~(0.10)$ & $0.509~(0.10)$ & $0.509~(0.10)$ & $0.502~(0.09)$ & $0.501~(0.10)$\\
					\hline
			\end{tabular}}
		\end{footnotesize}
	\end{center}
\end{table}


\section{Discussion and future work}
\label{sec:conclusion}

We have introduced a novel TFCCA approach for non-Euclidean functional data, and in particular, probability densities and shapes. Our approach is based on local linearizations of the data in tangent spaces and dimension reduction. The framework, in principle, can be used on several other nonlinear functional data objects, especially in scenarios when objects have been normalized to unit length to facilitate scale invariant analysis. In these cases, the relatively simple geometry of the Hilbert unit sphere can be used to substantial benefit. 

The main limitation of the proposed framework is exposed when datasets contain high variability and restricting attention to a single (or groupwise) tangent space(s) might result in distorted estimates of correlations. On the other hand, as seen in the analysis of the GBM tumor dataset, when variances are relatively small, the tangent space framework is quite useful. Thus, a natural extension of the framework is to develop intrinsic FCCA on manifolds that avoids any linearization and is able to uncover any potential nonlinear correlations. Although not directly related to FCCA, Dubey and M\"{u}ller \cite{https://doi.org/10.1111/rssb.12337} discuss intrinsic approaches for time varying functional data on general metric spaces based on the notion of metric covariance. We plan to explore whether this idea can be applied in our setting.

To summarize a GBM tumor's texture information, we use a univariate voxel value PDF. Incorporating spatial information of the voxels into the PDF would result in a richer representation of a tumor's appearance. However, the tumor regions for different GBM patients are not in direct correspondence with each other, making any subsequent comparisons difficult without first registering the tumor regions. As such, we leave this interesting research direction as future work. Finally, we assume that the voxel value PDFs are strictly positive. However, in some applications, one may encounter PDFs of interest with disconnected support. 

\section*{Acknowledgements}

We thank Arvind Rao (University of Michigan) for sharing the GBM tumor dataset. We also thank the Editor, Associate Editor and referees for their comments which led to an improved version of the manuscript. This research was partially funded by grants NSF DMS-1613054, NIH R37-CA214955 (to SK and KB),  NSF DMS-2015374 and EPSRC EP/V048104/1 (to KB), and NSF CCF-1740761, NSF DMS-2015226 and NSF CCF-1839252 (to SK).

\bibliographystyle{plain}
\bibliography{mybibfile}
\section{Supplementary material}
\section{Simulation study for shapes}
\label{sec:simul_shape}

We design a simulation study for tangent FCCA for shape curves. Here, we also investigate differences in estimated canonical correlations depending on whether the data is represented in a common tangent space versus in separate tangent spaces (option (i) in Section 3.2 in the main article). However, in this case, we do not use a common tangent space defined at the mean shape estimated from pooled data; this is due to registration inaccuracies that can be introduced while estimating the mean when the two groups are characterized by heterogeneous shapes. Instead, we use option (iii) as described in Section 3.2 in the main article. We parallel transport the tangent vector data and eigenbases for Group 1 from its tangent space (at the mean shape estimated using data in Group 1) to the tangent space at the mean shape estimated using data in Group 2. Then, we estimate the FPC coefficient matrices for each group in this common tangent space.  

\begin{figure}[!t]
	\begin{center}
		\begin{tabular}{|c|c|}
			\hline
			(a) Group 1 & (b) Group 2 \\
			\hline
			\begin{minipage}{.35\textwidth}
				\includegraphics[width=\linewidth]{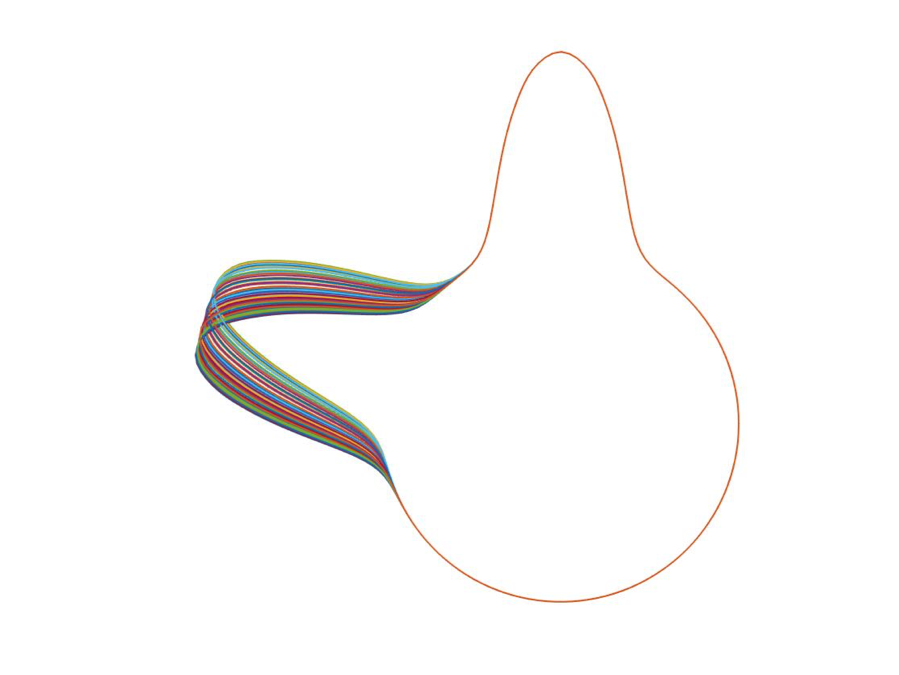}
			\end{minipage}
			&
			\begin{minipage}{.35\textwidth}
				\includegraphics[width=\linewidth]{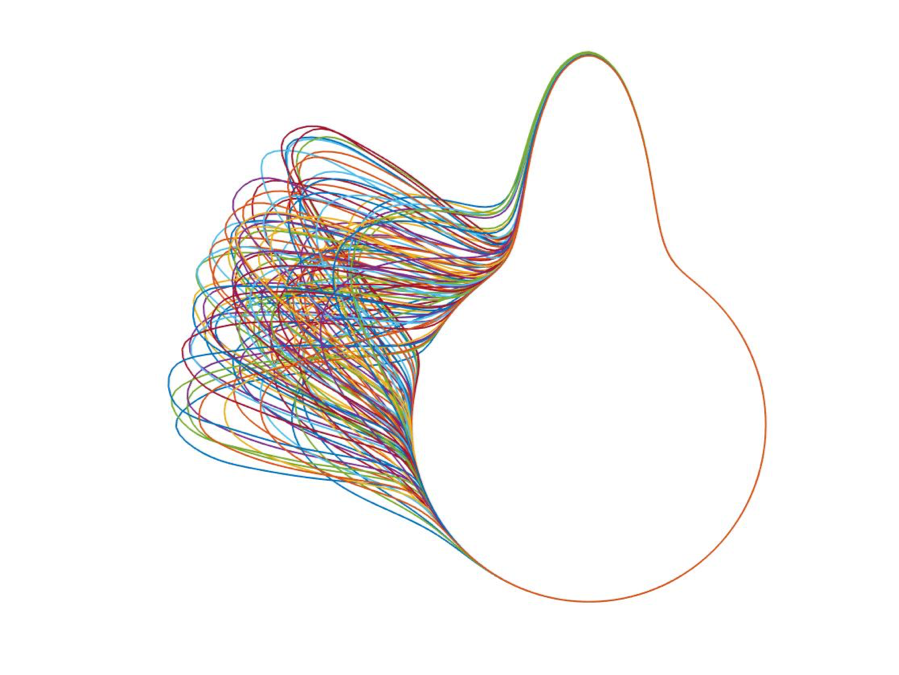}
			\end{minipage}
			\\
			\hline
		\end{tabular}
	\end{center}
	\caption{100 simulated closed curves from two different groups.}
	\label{fig:simshape}
\end{figure}

The planar curve data used in this simulation is shown in Figure \ref{fig:simshape}. The data in each case was generated by peaks of a von Mises probability density to a unit circle. One of the peaks is fixed across all of the data, i.e., all curves in Group 1 and Group 2 have a peak with the same shape pointing ``north''. In Group 1, we added a second peak that slightly varies in location; all of the second peaks in Group 1 have the same exact shape. In Group 2, we added a second peak that varies much more in location and also varies in shape (some peaks are thinner/thicker than others). 

\begin{table}[!t]
	\caption{Comparison of ``ground truth'' correlation and re-estimated canonical correlations using options (i) ($\hat{\bm \rho}_s$) and (iii) ($\hat{\bm \rho}_c$). In each case, we use $r=3$.}
	\label{tab:simshape}
	\begin{center}
		\begin{footnotesize}
			\makebox[\textwidth]{\begin{tabular}{c|c|c|c}
					\hline
					Setting & Highly Correlated & Moderately Correlated & Weakly Correlated \\ \hline
					$\bm \rho$ & $0.9934$ & $0.4815$ & $0.0533$ \\
					$\hat{\bm \rho}_s$ & $(0.9936,0.1673,0.0188)$ & $(0.6025,0.1124,0.0052)$ & $(0.1738,0.0905,0.0420)$ \\
					$\hat{\bm \rho}_c$ & $(0.9936,0.1671,0.0188)$ & $(0.6025,0.1123,0.0051)$ & $(0.1739,0.0905,0.0420)$ \\
					\hline
			\end{tabular}}
		\end{footnotesize}
	\end{center}
\end{table}
Note that, in general, it is difficult to design simulation studies to assess correlations across shape curves, since shape quantification requires complex optimization over nuisance variation (rotation and re-parameterization). Thus, as a proxy, we generate simulated data under three different scenarios: highly correlated, moderately correlated and weakly correlated location of the second peak across the two groups. We then quantify the ``ground truth'' correlation via the correlation between the locations of the second peak; the corresponding correlation values for the three scenarios are denoted by $\bm \rho$. We denote the canonical correlations estimated via FPC coefficient representations of the two groups in separate tangent spaces and a common tangent space as $\bm \rho_{s}$ and $\bm \rho_{c}$, respectively. Table \ref{tab:simshape} compares $\hat{\bm \rho}_c$ and $\hat{\bm \rho}_s$ to $\bm \rho$ based on the three different scenarios. The leading estimated canonical correlation in each case is larger than $\bm \rho$, although not by a large amount. This result is not unexpected, since the proposed TFCCA method captures correlations between entire shapes while the ``ground truth'' correlation is only related to the correlation in the location of the second peak. Furthermore, there appears to be very small differences between $\hat{\bm \rho}_s$ and $\hat{\bm \rho}_c$ in each scenario. 

\begin{figure}[!t]
	\begin{center}
		\begin{tabular}{|c|c|}
			\hline
			Group 1 & Group 2 \\
			\hline
			\begin{minipage}{.4\textwidth}
				\includegraphics[width=\linewidth]{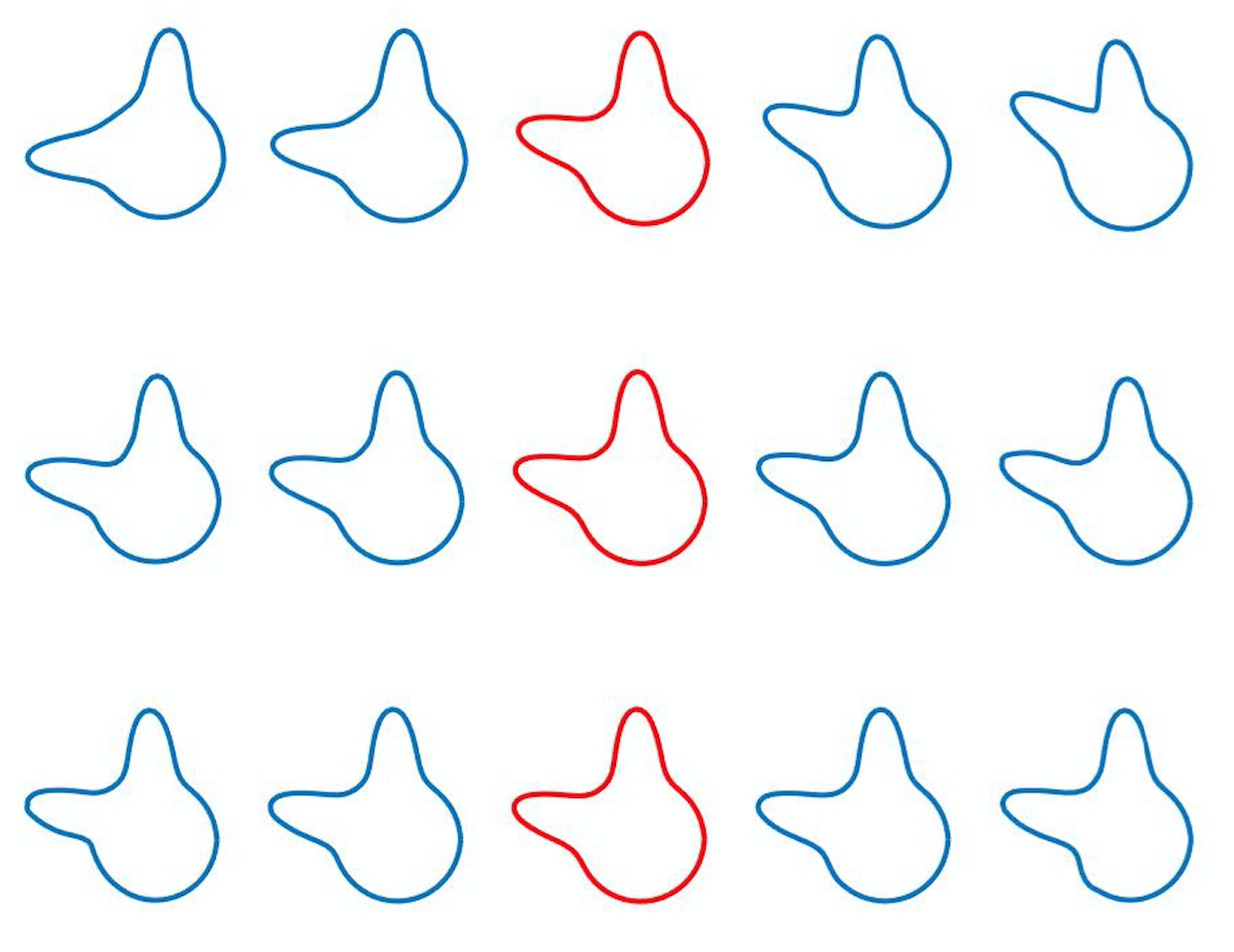}
			\end{minipage}
			&
			\begin{minipage}{.4\textwidth}
				\includegraphics[width=\linewidth]{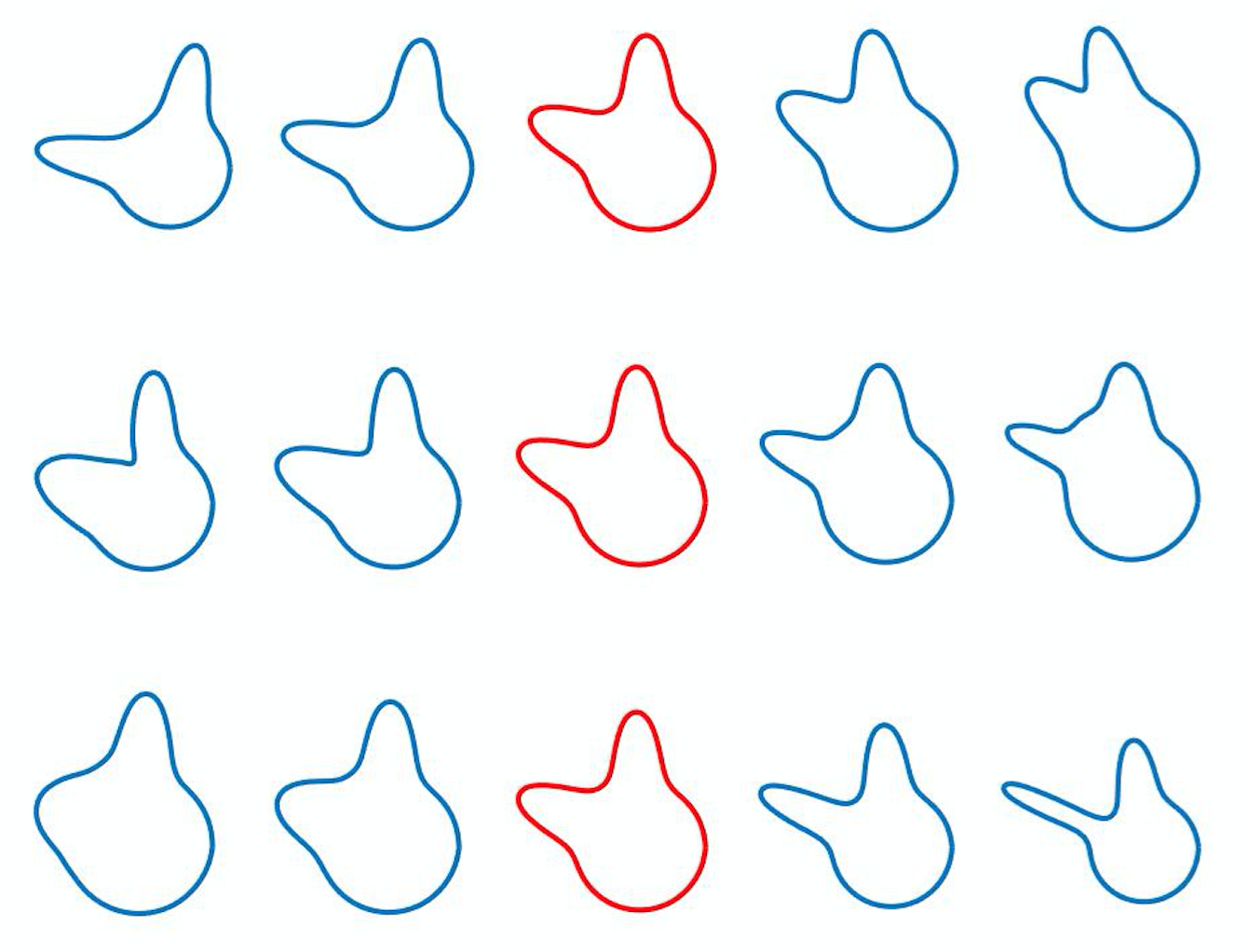}
			\end{minipage}
			\\
			\hline
		\end{tabular}
	\end{center}
	\caption{First three canonical variate directions (top to bottom) for Groups 1 and 2 estimated on separate tangent spaces under the highly correlated setting.}
	\label{fig:pccc1}
\end{figure}

\begin{figure}[!t]
	\begin{center}
		\begin{tabular}{|c|c|c|}
			\hline
			& PDFs & Shapes \\
			\hline
			T1
			&
			\begin{minipage}{.5\textwidth}
				\includegraphics[width=\linewidth]{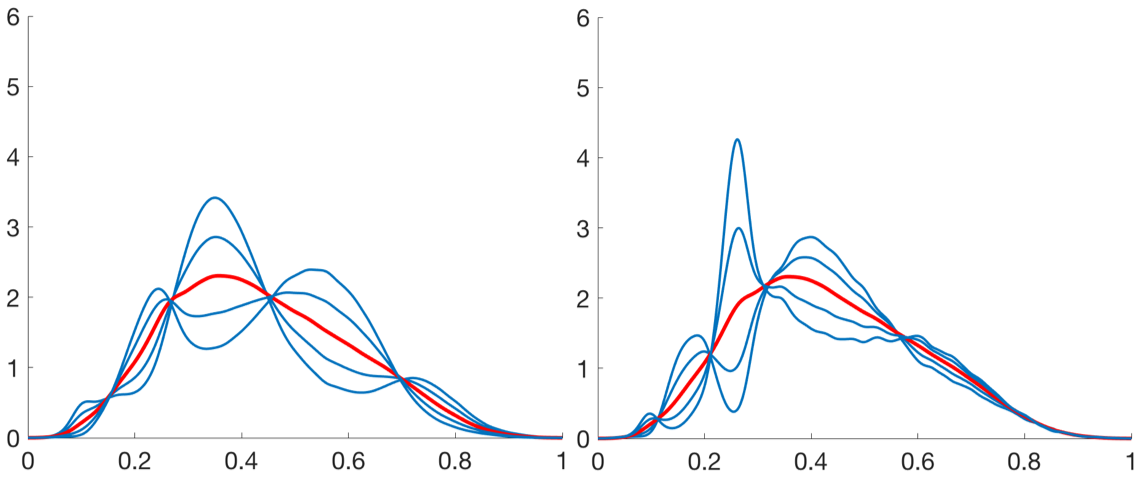}
			\end{minipage}
			&
			\begin{minipage}{.3\textwidth}
				\includegraphics[width=\linewidth]{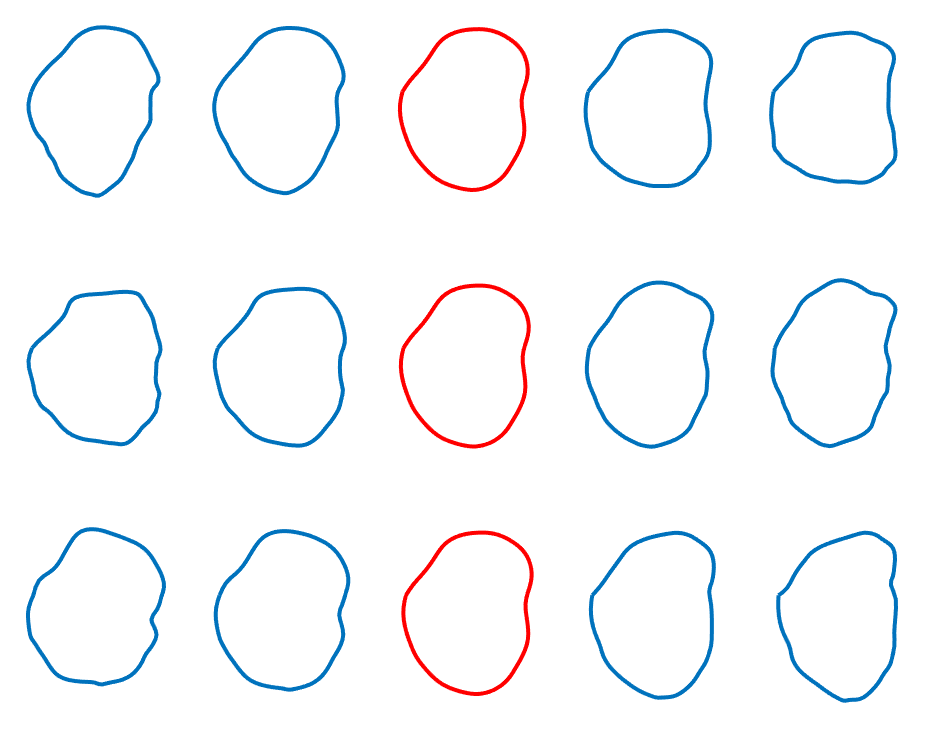}
			\end{minipage}
			\\
			\hline
			T2
			&
			\begin{minipage}{.5\textwidth}
				\includegraphics[width=\linewidth]{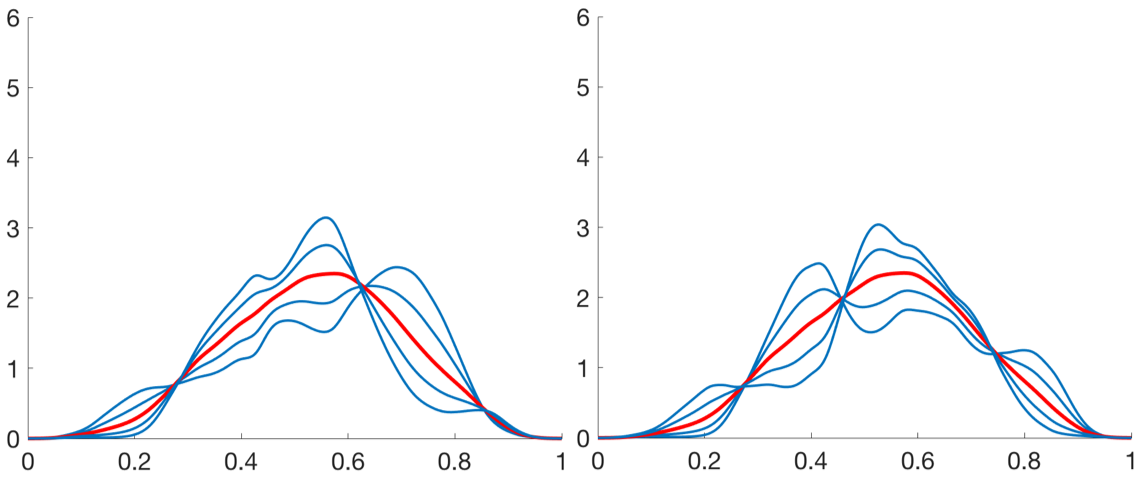}
			\end{minipage}
			&
			\begin{minipage}{.3\textwidth}
				\includegraphics[width=\linewidth]{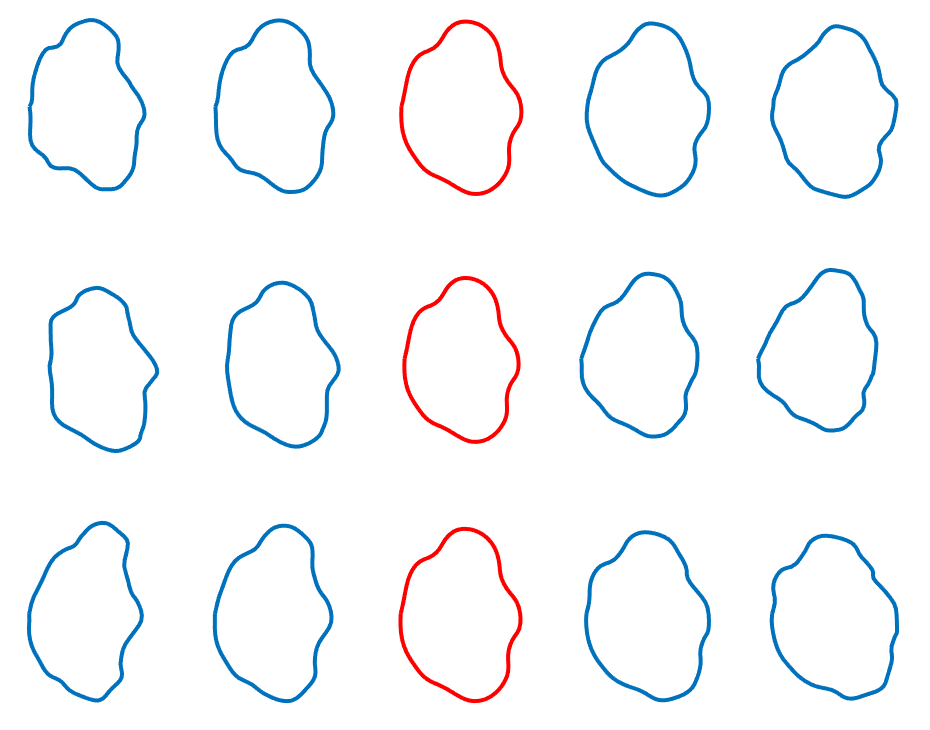}
			\end{minipage}
			\\
			\hline
		\end{tabular}
	\end{center}
	\caption{First two (left to right) and three (top to bottom) canonical variate directions for voxel value PDFs (left) and shapes (right), respectively, for the T1 top and T2 modalities.}
	\label{fig:ccc}
\end{figure}

Figure \ref{fig:pccc1} displays the leading three canonical variate directions for each group under the highly correlated setting. The first direction clearly captures the movement of the second peak in each group; the corresponding canonical correlation is very large at $0.9936$. The second and third directions appear to capture the shape differences in the second peak and how well the two peaks are separated. These features are more clearly observed in the canonical variate directions for Group 2. The corresponding canonical correlations are very small at 0.1673 and 0.0188, respectively. 

\begin{figure}[!t]
	\begin{center}
		\begin{tabular}{|c|}
			\hline
			T1 \\
			\hline
			\begin{minipage}{.9\textwidth}
				\includegraphics[width=\linewidth]{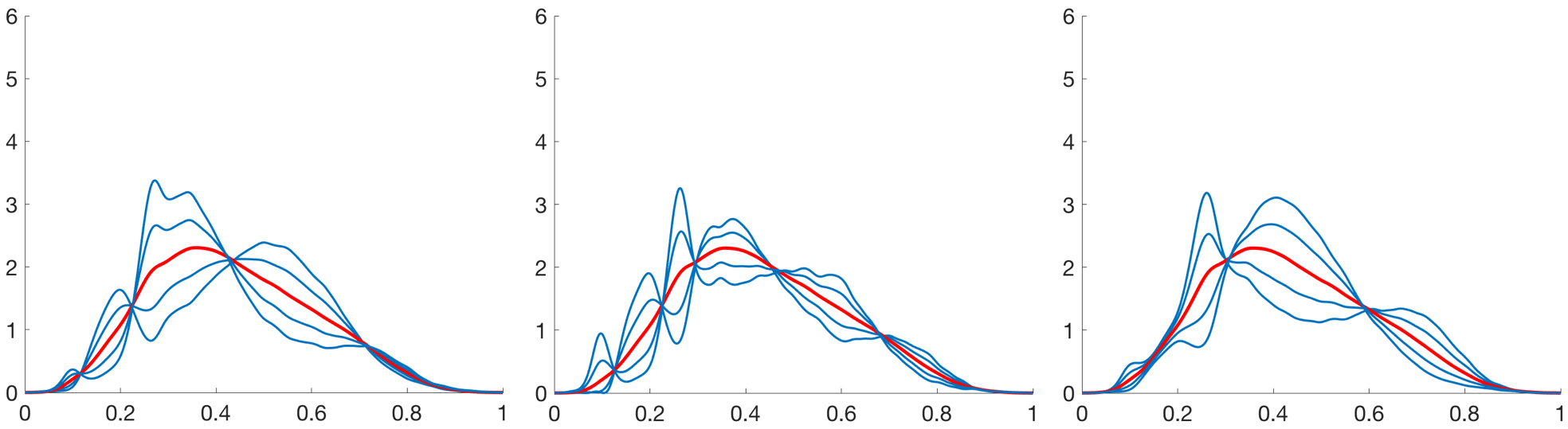}
			\end{minipage}
			\\
			\hline
			T2 \\
			\hline
			\begin{minipage}{.9\textwidth}
				\includegraphics[width=\linewidth]{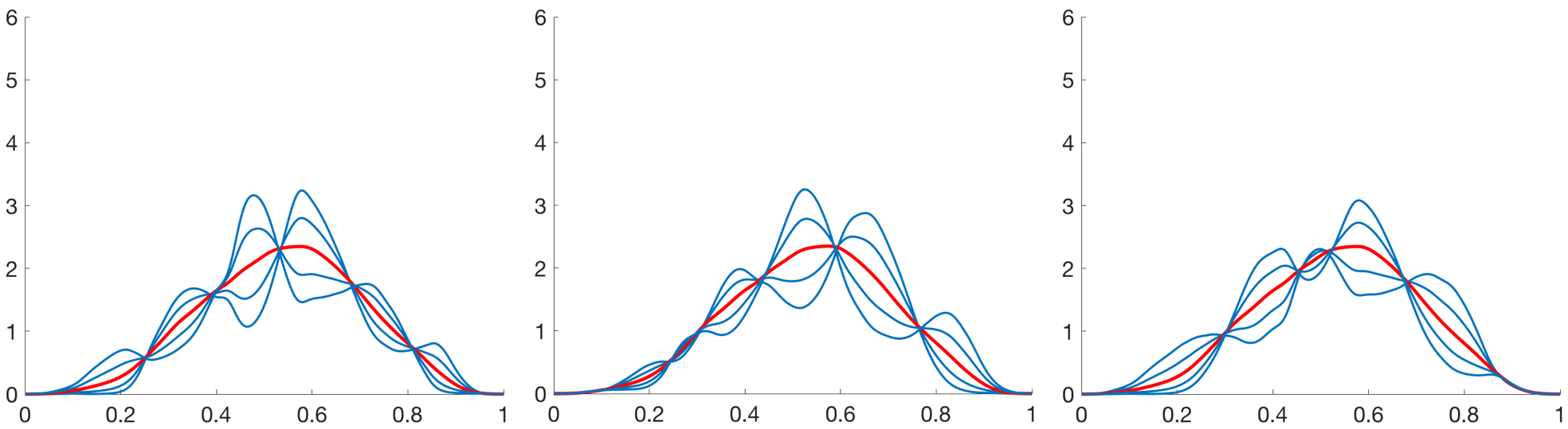}
			\end{minipage}
			\\
			\hline
		\end{tabular}
	\end{center}
	\caption{First three canonical variate directions (left to right) for PDFs from T1 and T2 modalities estimated via CVR.}
	\label{fig:cvr}
\end{figure}
\begin{figure}[!t]
	\begin{center}
		\begin{tabular}{|c|c|}
			\hline
			T1 & T2 \\
			\hline
			\begin{minipage}{.4\textwidth}
				\includegraphics[width=\linewidth]{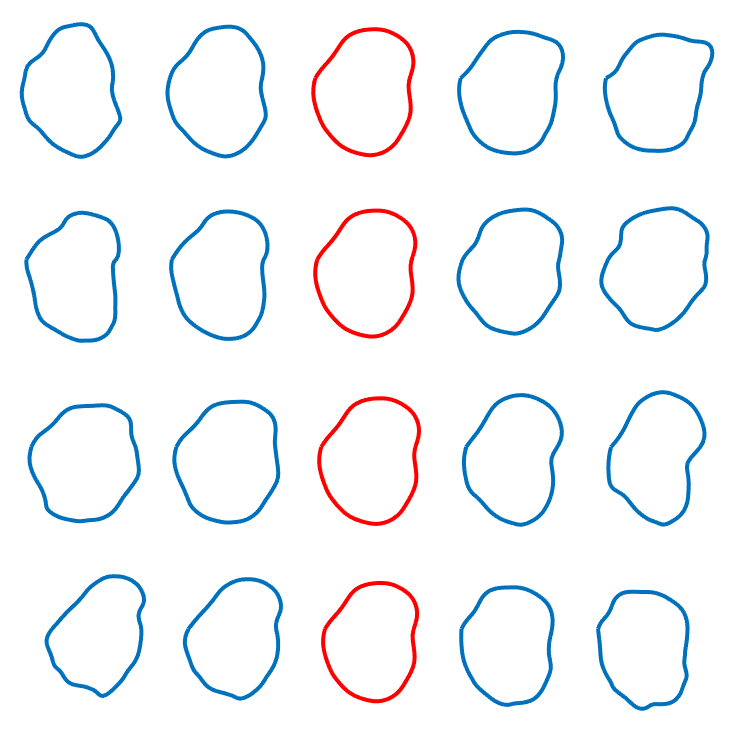}
			\end{minipage}
			&
			\begin{minipage}{.4\textwidth}
				\includegraphics[width=\linewidth]{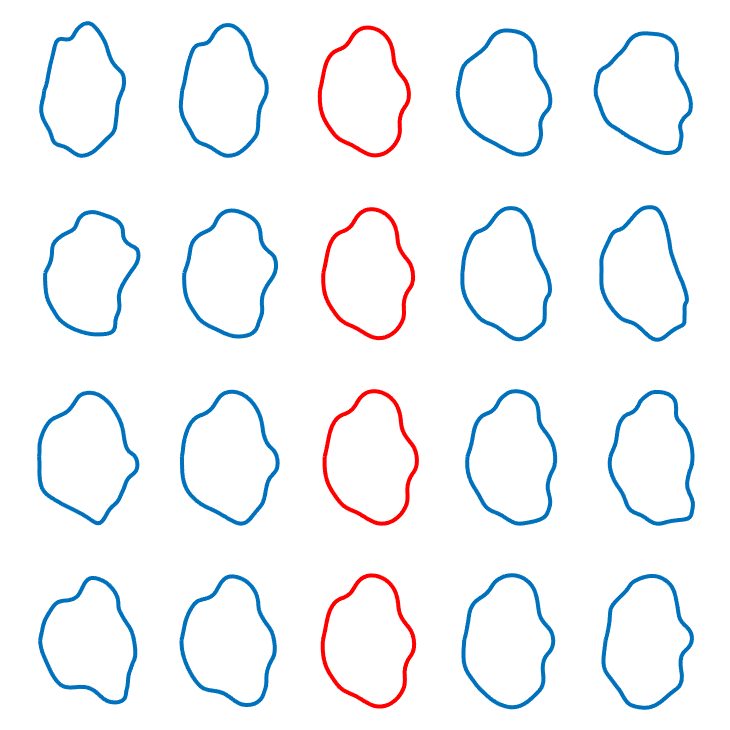}
			\end{minipage}
			\\
			\hline
		\end{tabular}
	\end{center}
	\caption{First four canonical variate directions (top to bottom) for GBM tumor shapes from T1 and T2 modalities estimated via CVR.}
	\label{fig:pccc3}
\end{figure}

\section{CCA for PDFs and shapes estimated from the same MRI modality}
The proposed TFCCA approach can also be used to study associations between textural features of the tumors (via voxel value PDFs) and their shapes within the same MRI modality. In this case, for the T1 modality, we use five and 24 FPC coefficients for PDFs and shapes, respectively, and for the T2 modality, we use five and 28 PC coefficients, respectively. In all cases, the corresponding FPCs explain at least $95\%$ of the total variation in the data. We apply the proposed TFCCA approach to PDFs and shapes from each modality separately to explore canonical variate directions that capture the most correlation between textural and shape features of the GBM tumors. The estimated canonical correlations between PDFs and shapes for the T1 and T2 modalities are $(0.7907,~0.7059,~0.6275,~0.3922,~0.3843)$ and $(0.8484,~0.7688,~0.7213,~0.6815,~0.5327)$, respectively. The large leading canonical correlations indicate that textural features of the tumors are indeed associated with their shape variation in both modalities. It also appears that associations are stronger in the T2 modality. The corresponding first two canonical variate directions for PDFs and first three canonical variate directions for shapes, estimated for each modality, are displayed in Figure \ref{fig:ccc}. As expected, these directions are quite different from the ones displayed in Figures 4 and 5 in the main article. These directions capture the variability in voxel value PDFs and shapes that are most correlated within each modality. 


\section{Canonical variate regression directions}

In Figures \ref{fig:cvr} and \ref{fig:pccc3}, we display the CVR-based canonical variate directions for PDFs and shapes, respectively. These directions are slightly different from those estimated using TFCCA and displayed in Figures 4 and 5 in the main article, as they place more weight on predicting log-survival than maximizing correlation. 
\end{document}